\documentclass[prb,reprint,10pt,aps]{revtex4-1}
\usepackage{amsmath,amssymb}
\usepackage{upgreek}
\usepackage{comment}
\usepackage{color}
\usepackage{graphicx}
\usepackage{multirow}
\usepackage{here}

\begin{document}
	
	\title{Intrinsic and extrinsic effects on intraband optical conductivity of hot carriers in photoexcited graphene }%
	
	\author{Masatsugu Yamashita}%
	\email[e-mail: ]{m-yama@riken.jp}
	\author{Chiko Otani}%
	\affiliation{Terahertz Sensing and Imaging Team, RIKEN Center for Advanced Photonics, RIKEN, 519-1399 Aramaki-Aoba Aoba-ku, Sendai, Miyagi, 980-0845, Japan}
	\date{12, January 2021}%
	
	\begin{abstract}
		We present a numerical study on the intraband optical conductivity of hot carriers at quasi-equilibria in photoexcited graphene based on the semiclassical Boltzmann transport equations (BTE) with the aim of understanding the effects of intrinsic optical phonon and extrinsic coulomb scattering caused by charged impurities at the graphene--substrate 
		interface. Instead of using full-BTE solutions, we employ iterative solutions of the BTE and the comprehensive model for the temporal evolutions of hot-carrier temperature and 
		hot-optical-phonon occupations to reduce computational costs. Undoped graphene exhibits large positive photoconductivity owing to the increase in thermally excited carriers 
		and the reduction in charged impurity scattering. The frequency dependencies of the photoconductivity in undoped graphene having high concentrations of charged impurities significantly deviate from those observed in the simple Drude model, which can be attributed to temporally varying charged impurity scattering during terahertz (THz) probing in the hot-carrier cooling process. Heavily doped graphene exhibits small negative photoconductivity similar to that of the Drude model. In this case, charged impurity scattering is substantially suppressed by the carrier-screening effect, and the temperature dependencies of the Drude weight and optical phonon scattering governs the negative photoconductivity. In lightly doped graphene, the appearance of negative and positive photoconductivity depends on the frequency and the crossover from negative photoconductivity to positive emerges from increasing the charged impurity concentration. This indicates the change of the dominant scattering mechanism from optical phonons to charged impurities. Moreover, the photoconductivity spectra depend not only on the material property of the graphene sample but also on the waveform of the THz-probe pulse. Our approach provides a quantitative understanding of non-Drude behaviors and the temporal evolution of photoconductivity in graphene, which is useful for understanding hot carrier behavior and supports the development of future graphene opto-electronic devices.
	\end{abstract}
	
	\maketitle
	
	
	\section{Introduction}
	In many optoelectronic graphene applications (e.g., photodetection \cite{Koppens2014a}, plasmonics \cite{Grigorenko2012}, light harvesting \cite{Bonaccorso2015a}, data communication \cite{Xia2009, Schall2014}, ultrafast laser \cite{Bao2009a, Sun2010, Yoshikawa2017a}, and terahertz (THz) technologies \cite{ Lee2012a,Sensale-rodriguez2012a,Cai2014a, Alonso-gonzalez2017, Hafez2018a}), it is crucial to understand the carrier dynamics that occur following photoexcitation and their influence on electrical and optical conductivities. The ultrafast dynamics of hot carriers in graphene has been studied intensively using various ultrafast spectroscopic techniques \cite{Sun2008a, Lui2010b, Mihnev2016a, Tomadin2018c, Tani2012c, Breusing2011a, Hale2011a,Brida2013, Tomadin2013b, Mittendorff2014a, Gierz2013c, Docherty2012e, Frenzel2014d} to understand the fundamental carrier--carrier scattering and carrier--phonon relaxation processes of the 2D massless Dirac fermion (MDF). Numerous experiments using optical-pump THz-probe spectroscopy (OPTP) \cite{Mihnev2016a, Tomadin2018c, Docherty2012e, Frenzel2014d, Tielrooij2013a, Shi2014a, Jensen2014b, Kar2014a, Heyman2015a, George2008a, Strait2011a, Boubanga-Tombet2012a,Frenzel2013a, Jnawali2013b, Lin2013d} have revealed  unusual behaviors of graphene undergoing positive and negative changes of intraband optical conductivity. However, these results were interpreted using the framework of the phenomenological model for a near-equilibrium condition \cite{Docherty2012e, Frenzel2014d, Shi2014a, Heyman2015a, Jnawali2013b}. Positive photoinduced THz conductivity has been explained as an enhanced free-carrier intraband absorption that occurs upon photoexcitation. In contrast, negative photoinduced THz conductivity has been variously ascribed to stimulated THz emission \cite{Docherty2012e}, increased carrier scattering with optical phonons or charge impurities \cite{Frenzel2013a,Jnawali2013b}, and carrier heating \cite{Frenzel2014d, Tielrooij2013a, Shi2014a, Jensen2014b}. 
	
	Microscopic models based on semiconductor Bloch equations for graphene have been employed for the quantitative and qualitative understanding of hot-carrier dynamics considering intrinsic effects, such as carrier--carrier and carrier--phonon scattering \cite{Malic2011c, Malic2012a, Mihnev2016a}. However, in addition to the intrinsic effects, extrinsic effects, such as charged impurities, surface optical phonons, and dielectric properties on the substrate, play essential roles in the electrical and optical conductivities of hot carriers \cite{Tomadin2018c, Frenzel2014d, Shi2014a}. Furthermore, most previous studies discussed hot-carrier dynamics based on transient transmission change, which determines the response around the center frequencies of the THz-probe pulses. To obtain more quantitative insights into hot-carrier dynamics governed by intrinsic and extrinsic effects, analyses based on the frequency dependence of THz conductivity are necessary. Such effects are essential for understanding the physics underlying various graphene-based device applications. However, it remains a practical numerical challenge to obtain the frequency-dependent optical conductivity for intraband and interband transitions with the full solution of a carrier distribution function by solving the semiconductor Bloch equation in the 2D momentum space of graphene, even if only intrinsic carrier--carrier and carrier--optical phonon scatterings are considered \cite{Mihnev2016a}. Therefore, a suitably approximate method is required for understanding the hot-carrier dynamics affected by extrinsic effects.
	
	In this study, we calculate the frequency-dependent intraband optical conductivity of hot carriers in photoexcited graphene based on the Boltzmann transport equation (BTE), including the intrinsic and extrinsic interactions in the collision term. Because the intraband transition is dominant in the THz-frequency region of 0.1--10 THz, the microscopic polarization for interband transition in the semiconductor Bloch equation may be negligible in the calculation of optical conductivity at the quasi-equilibrium carrier distribution in graphene after photoexcitation \cite{Mihnev2016a, Tani2012c, Tomadin2013b}. Under such conditions, the semiclassical BTE may offer an alternative \cite{Ridley2013a, Haug2009a}. Under linearly polarized photoexcitations, highly nonequilibrium anisotropic electron and hole distributions are created and rapidly relaxed to the uniform hot Fermi--Dirac (FD) distribution having two different chemical potentials in the conduction and valence bands via carrier--carrier and carrier--optical phonon scattering within several tens of femtoseconds \cite{Mittendorff2014a, Malic2011c, Malic2012a}. Following the recombination of the photoexcited electron and hole pairs via Auger recombination and the interband optical phonon-emission process, the carrier distribution is relaxed to the hot-FD distribution with a single chemical potential \cite{Gierz2013c}. Thereafter, the thermalized carriers and optical phonons at quasi-equilibrium cool to equilibrium via the energy transfer from the hot carriers and optical phonons to other types of phonons through phonon--phonon interactions caused by lattice anharmonicity \cite{Wu2012a} and supercollision (SC) carrier-cooling, which produce disorder-mediated electron–acoustic phonon scatterings \cite{Song2012a, Feve2012a, Someya2017b}. Because the energy relaxation caused by the optical and acoustic phonon emission is inefficient for low-energy carriers near the Dirac point, the SC carrier-cooling process becomes important. To consider the cooling processes of carriers and optical phonon modes, at least five coupled BTEs expressed as nonlinear integrodifferential equations must be solved for carriers in conduction and valence bands and in three dominant optical phonon modes \cite{Piscanec2004b} of graphene. Further reductions in computational costs are still required.

	The BTE solution using relaxation-time approximation (RTA) has been used extensively, because it makes it easy to solve BTE \cite{Lundstrom2009a} (see Appendix B). The RTA is valid in the spherical energy band for elastic scattering under low-field conditions. For the inelastic scattering process, RTA is only valid for isotropic scattering in non-degenerate semiconductors, where the FD distribution can be approximated using the Boltzmann distribution. However, in the case of graphene, the RTA solution underestimates or overestimates the relaxation rate for the inelastic scattering, because the Boltzmann distribution is not valid. Furthermore, the sub-picosecond temporal resolution of the OPTP experiments cannot instantaneously capture the carrier dynamics. The carrier distribution and momentum relaxation rate in the cooling process change significantly within a probe time that is approximately equal to the THz-pulse duration. Therefore, the THz conductivity cannot be adequately analyzed by the calculation based on the RTA. To ensure calculation accuracy, we employ an iterative method \cite{Rees1969, Rode1971a, Willardson1972a} that provides the BTE solution for the intraband complex conductivity in graphene near the (quasi-)equilibrium under a weak electric field with appropriate accuracy. Moreover, to calculate the hot-carrier intraband conductivity of graphene, we perform analysis using the BTE combined with a comprehensive temperature model based on rate equations to describe the temporal evolution of hot-carrier temperature and hot-optical-phonon occupations \cite{Lui2010b, Lin2013d, Wu2012a, Someya2017b, Wang2010a}. We present numerical simulations of the intraband optical conductivity of hot carriers in photoexcited graphene with different Fermi energies, considering the intrinsic and extrinsic carrier scatterings and the temporal variations of the carrier distribution and momentum relaxation rate during THz probing.
	
	The remainder of this paper is organized as follows. In Section II, we present the iterative solutions of the BTEs for graphene at near-equilibrium under a weak electric field to calculate the direct current (DC) and intraband optical conductivity. Furthermore, a temperature model based on coupled rate equations is introduced to apply the iterative solutions to the quasi-equilibrium hot-carrier state in the cooling process following photoexcitation. Section III presents our numerical results for the THz conductivity of undoped and heavily doped graphene. Finally, we discuss the origin of unique behavior of lightly doped graphene, and the major conclusions are summarized in Section IV.

	\section{Intraband optical conductivity calculation of graphene}
	\subsection*{A. Iterative solutions of BTE in steady state}
	
	The iterative solution of the BTE for obtaining the steady-state conductivity of graphene was introduced in Ref.\,\cite{Willardson1972a}. The BTE in a homogenous system under a time-dependent electric field describes the temporal evolution of the carrier distribution \cite{Lundstrom2009a, Hamaguchi2010a}, which is given as 
		\begin{equation}
	\frac{\partial f_{\lambda}(\boldsymbol{k}, t)}{\partial t}=-\frac{(-e)}{\hbar} \boldsymbol{E}(t) \frac{\partial f_{\lambda}(\boldsymbol{k}, t)}{\partial \boldsymbol{k}}+\left.\frac{\partial f_{\lambda}(\boldsymbol{k}, t)}{\partial t}\right|_{\text {collision }}.
	\end{equation}
	Here, $f_{\mathit{\lambda}}(\boldsymbol{k}, t)$ is the electron distribution function for the conduction band ($\lambda = 1$) and valence band ($\lambda =  -1$). $\boldsymbol{k}$ is the wave vector of the carriers, $e$ is the elementary charge, $\boldsymbol{E}(t)$ is the electric field, and $\partial f_{\lambda}(\boldsymbol{k}, t) /\left.\partial t\right|_{\text {collision }}$ is the collision term describing the change in the distribution function via carrier scattering. The BTE in a steady state under a constant electric field, $\textbf{\textit{E}}$ ($d f_{\lambda}(\boldsymbol{k}, t) / d t = 0$), is given by
		\begin{equation}
	\frac{(-e)}{\hbar} \boldsymbol{E} \frac{\partial f_{\lambda}(\boldsymbol{k})}{\partial \boldsymbol{k}}=\left.\frac{\partial f_{\lambda}(\boldsymbol{k})}{\partial t}\right|_{\text {collision }}.
	\end{equation}
		For spherical bands under a low field, $\boldsymbol{E}$, the general solution of Eq.\;(2) is provided approximately by the first two terms of the zone spherical expansion.
		\begin{equation}
	f_{\lambda}(\boldsymbol{k})=f_{0}\left(\varepsilon_{\lambda \boldsymbol{k}}\right)+g\left(\varepsilon_{\lambda \boldsymbol{k}}\right) \cos \alpha.
	\end{equation}
	Here, $f_{0}\left(\varepsilon_{\lambda \boldsymbol{k}}\right)=1 /\left[\exp \left\{\left(\varepsilon_{\lambda \boldsymbol{k}}-\mu\left(T_{\mathrm{e}}\right)\right) / k_{\mathrm{B}} T_{\mathrm{e}}\right\}+1\right]$ is the FD distribution for the corresponding equilibrium electron distribution at the electron temperature, $T_{\mathrm{e}}$, $\varepsilon_{\lambda \boldsymbol{k}}=\pm \hbar v_{\mathrm{F}}|\boldsymbol{k}|$ ($\varepsilon_{1 \boldsymbol{k}} \geq 0$, for the conduction band. $\varepsilon_{-1 \boldsymbol{k}} \leq 0$ for the valence band) is the electron energy within the Dirac approximation of the graphene energy-band structure \cite{Novoselov2009b}, and $\mu\left(T_{\mathrm{e}}\right)$ is the temperature-dependent chemical potential of the 2D-MDF (see Appendix A and Ref.\,\cite{Ando2006a, Hwang2009e, Frenzel2014d}). Moreover, $g\left(\varepsilon_{\lambda \boldsymbol{k}}\right)$ is the perturbation part of the distribution, and $\alpha$ is the angle between $\boldsymbol{E}$ and $\boldsymbol{k}$. 
	
	We consider the collision term, 
	\begin{equation}
	\frac{\partial f_{\lambda}(\boldsymbol{k})}{\partial t}|_{\text {collision }}=\sum_{\eta, \lambda^{\prime}} C_{\lambda \lambda^{\prime}}^{\eta}(\boldsymbol{k})+C_{\lambda}^{\mathrm{el}}(\boldsymbol{k}),\end{equation}
	while accounting for the scattering of electrons having different optical phonon modes, $\eta$, in $C_{\lambda \lambda}^{\eta}$, including both intraband $\left(\lambda=\lambda^{\prime}\right)$ and interband $\left(\lambda \neq \lambda^{\prime}\right)$ processes with the elastic scattering processes in $C_{\lambda}^{\mathrm{el}}(\boldsymbol{k}).$ The carrier collision 
	term, $C_{\lambda \lambda^{\prime}}^{\eta}(\boldsymbol{k})$, for the interaction of the electron and optical phonons is expressed as
	\begin{equation}\begin{aligned}
	C_{\lambda \lambda^{\prime}}^{\eta}(\boldsymbol{k}) =\sum_{k^{\prime}}&\left\{P_{\boldsymbol{k}^{\prime} \lambda^{\prime} \boldsymbol{k} \lambda}^{\eta} f_{\lambda^{\prime}}\left(\boldsymbol{k}^{\prime}\right)\left(1-f_{\lambda}(\boldsymbol{k})\right)\right.\\
	&\left.-P_{\boldsymbol{k} \lambda \boldsymbol{k}^{\prime} \lambda^{\prime}}^{\eta} f_{\lambda}(\boldsymbol{k})\left(1-f_{\lambda^{\prime}}\left(\boldsymbol{k}^{\prime}\right)\right)\right\},
	\end{aligned}\end{equation}
	where $P_{\boldsymbol{k}^{\prime} \lambda^{\prime} \boldsymbol{k} \lambda}^{\eta}$ and $P_{\boldsymbol{k} \lambda \boldsymbol{k}^{\prime} \lambda^{\prime}}^{\eta}$ are the carrier scattering rates obtained by the optical phonon modes, $\eta$, between states $(\boldsymbol{k}^{\prime}, \lambda^{\prime}) \rightarrow(\boldsymbol{k}, \lambda)$ and
	$(\boldsymbol{k}, \lambda) \rightarrow(\boldsymbol{k}^{\prime}, \lambda^{\prime}),$ respectively. This is expressed by
	\begin{equation}
	P_{\boldsymbol{k} \lambda \boldsymbol{k}^{\prime} \lambda^{\prime}}^{\eta}=P_{\boldsymbol{k} \lambda \boldsymbol{k}^{\prime} \lambda^{\prime}}^{\mathrm{EM}, \eta}+P_{\boldsymbol{k} \lambda \boldsymbol{k}^{\prime} \lambda^{\prime}}^{\mathrm{AB}, \eta},
	\end{equation}
	which accounts for the phonon emission and absorption given by
	\begin{equation}\begin{aligned}
	P_{\boldsymbol{k} \lambda \boldsymbol{k}^{\prime} \lambda^{\prime}}^{\mathrm{EM/AB}, \eta} =&\frac{\pi\left|D_{\boldsymbol{k} \boldsymbol{k}^{\prime}}^{\eta}\right|^{2}}{\rho \omega_{\eta}}\left(n_{\eta}+\frac{1}{2} \pm \frac{1}{2}\right) \\
	& \times \delta\left(\varepsilon_{\lambda \boldsymbol{k}}-\varepsilon_{\lambda^{\prime} \boldsymbol{k}^{\prime}} \mp \hbar \omega_{\eta}\right) \delta\left(\boldsymbol{k}-\boldsymbol{k}^{\prime} \mp \boldsymbol{q}\right),
	\end{aligned}\end{equation}
	where $\left|D_{\boldsymbol{k} \boldsymbol{k}^{\prime}}^{\eta}\right|$ is the electron--phonon coupling $(\mathrm{EPC})$-matrix element defined by Ref.\,\cite{Piscanec2004b} (see APPENDIX B). $\rho = 7.6 \times 10^{-7}\,\mathrm{kgm}^{-2}$ is the area density of graphene, and $\omega_{\eta}$ and $n_{\eta}$ are the angular frequency and occupation of the optical phonons, respectively. Density functional theory (DFT) calculations have demonstrated that only three optical phonon modes contribute significantly to the inelastic scattering of electrons in graphene \cite{Piscanec2007c}. The first two relevant modes are longitudinal optical (LO) and transversal optical (TO) phonons near the $\mathrm{\Gamma}$ point with energies of $196\,\mathrm{meV}$ \cite{Lichtenberger2011c}, which contribute to intravalley scattering. Moreover, zone-boundary phonons have $\hbar \omega_{\mathrm{\bf{K}}}=161\,\mathrm{meV}$ \cite{Lichtenberger2011c} close to the $\mathbf{K}$ point and are responsible for the intervalley scattering process. The carrier-scattering rates obtained by the optical phonons in Eq.\;(7) account for phonon emission and absorption. The delta functions, $\delta\left(\varepsilon_{\lambda \boldsymbol{k}}-\varepsilon_{\lambda^{\prime} \boldsymbol{k}^{\prime}} \mp \hbar \omega_{\eta}\right)$ and $\delta\left(\boldsymbol{k}-\boldsymbol{k}^{\prime} \mp \boldsymbol{q}\right)$, in Eq.\;(7) arise from Fermi’s golden rule, thereby ensuring the conservation of energy and momentum, respectively. The EPC-matrix elements, $|D_{\boldsymbol{k k}^{\prime}}^{\eta}|^{2}$, for $\Gamma\mathchar`-\mathrm{L} \mathrm{O}, \Gamma\mathchar`-\mathrm{T} \mathrm{O},$ and $\mathbf{K}$ phonons are expressed by \cite{Piscanec2004b, Piscanec2007c}
	\begin{equation}
	\begin{aligned}
	\left|D_{\boldsymbol{k k}^{\prime}}^{\mathrm{\Gamma\mathchar`-LO} / \mathrm{TO}}\right|^{2}&=\left\langle D_{\Gamma}^{2}\right\rangle_{\mathrm{F}}\left\{1 \pm \cos \left(\theta+\theta^{\prime}\right)\right\}, \\
	\left|D_{\boldsymbol{k k}^{\prime}}^{\mathrm{\textbf{K}}}\right|^{2}&=\left\langle D_{\mathrm{\textbf{K}}}^{2}\right\rangle_{\mathrm{F}}\left\{1 \pm \cos \theta^{\prime\prime}\right\}.
	\end{aligned}
	\end{equation}
	Here, $\theta$ denotes the angle between $\boldsymbol{k}$ and $\boldsymbol{k^{\prime}-k}$, $\theta^{\prime}$ denotes the angle between $\boldsymbol{k}^{\prime}$ and $\boldsymbol{k^{\prime}-k}$, and $\theta^{\prime\prime}$ denotes the angle between $\boldsymbol{k}$ and $\boldsymbol{k^{\prime}}$. In the case of $\Gamma\mathchar`-\mathrm{LO}$ and $\mathrm{\bf{K}}$ phonons, the plus sign refers to interband processes, and for $\Gamma$-TO phonons, it refers to intraband processes. The EPC coefficients were obtained via DFT calculations \cite{Lazzeri2005d}. Their values are
	$\langle D_{\Gamma}^{2}\rangle_{\mathrm{F}}=45.6\,(\mathrm{eV} / \mathrm{\AA})^{2}$ and $\langle D_{\mathrm{\bf{K}}}^{2}\rangle_{\mathrm{F}}=92.1\,(\mathrm{eV} / \mathrm{\AA})^{2}$. Note that the value of the EPC coefficient, $\langle D_{\mathrm{\mathrm{\textbf{K}}}}^{2}\rangle_{\mathrm{F}}$, has been debated, owing to the renormalization effect resulting from electron--electron interaction \cite{Basko2009c, Ferrari2006b, Lazzeri2008b, Berciaud2009b, Gruneis2009b}. In the numerical calculations, we consider the DFT value for simplicity. The elastic term is given by
	\begin{equation}\begin{aligned}
	C_{\lambda}^{\mathrm{el}}(\boldsymbol{k})=\sum_{s}&\left\{P_{\boldsymbol{k}^{\prime} \textbf{\textit{k}}}^{s} f_{\lambda}\left(\boldsymbol{k}^{\prime}\right)\left(1-f_{\lambda}(\boldsymbol{k})\right)\right.\\
	&\left.-P_{\textbf{\textit{k}} \textbf{\textit{k}}^{\prime}}^{s} f_{\lambda}(\boldsymbol{k})\left(1-f_{\lambda}\left(\boldsymbol{k}^{\prime}\right)\right)\right\},
	\end{aligned}\end{equation}
	where $P_{\textbf{\textit{k}}^{\prime} \textbf{\textit{k}}}^{s}$ and $P_{\textbf{\textit{k}} \textbf{\textit{k}}^{\prime}}^{s}$ are the scattering rates for the elastic scatterings. The index, $s$, refers to the different elastic or quasi-elastic scattering modes, such as charged impurities and acoustic phonons. Substituting Eqs.
	(5) and (9) into Eq.\;(2), we get
	\begin{equation}
	\frac{(-e)E}{\hbar} \frac{\partial f_{\lambda}\left(\textbf{\textit{k}}\right)}{\partial k}=S_{\lambda}^{\mathrm{in}}-g\left(\varepsilon_{\lambda \textbf{\textit{k}}}\right)\left(S_{\lambda}^{\mathrm{out}}+\nu^{\mathrm{e} 1}\right),
	\end{equation}
	where $E=|\textbf{\textit{E}}|$ and $\textit{k}=|\textbf{\textit{k}}|$ are the magnitudes of the electric field and wavevector, respectively.
	\begin{equation}\begin{aligned}
	S_{\lambda}^{\mathrm{in}}=\sum_{\eta, \textbf{\textit{k}}^{\prime}, \lambda^{\prime}}& g\left(\varepsilon_{\lambda^{\prime} \textbf{\textit{k}}^{\prime}}\right) \cos \alpha\\
	\times& \left\{P_{\textbf{\textit{k}}^{\prime} \lambda^{\prime} \textbf{\textit{k}} \lambda}^{\eta}\left(1-f_{0}\left(\varepsilon_{\lambda^{\prime} \textbf{\textit{k}}}\right)\right)+P_{\textbf{\textit{k}} \lambda \textbf{\textit{k}}^{\prime} \lambda^{\prime}}^{\eta} f_{0}\left(\varepsilon_{\lambda^{\prime} \textbf{\textit{k}}}\right)\right\},\end{aligned}\end{equation}
	\begin{equation}\begin{aligned}
	S_{\lambda}^{\text{out}}=\sum_{\eta, \textbf{\textit{k}}^{\prime} \lambda^{\prime}}\left\{P_{\textbf{\textit{k}} \lambda \textbf{\textit{k}}^{\prime} \lambda^{\prime}}^{\eta}\left(1-f_{0}\left(\varepsilon_{\lambda^{\prime} \textbf{\textit{k}}^{\prime}}\right)\right)+P_{\textbf{\textit{k}}^{\prime} \lambda^{\prime} \textbf{\textit{k}} \lambda}^{\eta} f_{0}\left(\varepsilon_{\lambda^{\prime} \textbf{\textit{k}}^{\prime}}\right)\right\}
	\end{aligned} \end{equation}
	are the net in- and out-scattering rates for inelastic scattering, respectively. Furthermore,
	\begin{equation}
	\nu^{\mathrm{el}}=\sum_{s} \nu^{s}=\sum_{s} \int d \boldsymbol{k}^{\prime}\left(1-\cos \theta^{\prime\prime} \right) P_{\textbf{\textit{k}} \textbf{\textit{k}}^{\prime}}^{s}
	\end{equation}
	is the total relaxation rate, which is the summation of the relaxation rate, $\nu^{s}$, caused by the different elastic-scattering processes indicated by $s$ using RTA. We consider the elastic carrier scattering via charged impurities as weak scatterers and acoustic phonons (for the formula of $\nu^{\text {el}},$ see Appendix B). Using the contraction mapping principle \cite{Rall1969a}, Eq.\;(10) is numerically solved using an iterative procedure for the given inelastic and elastic scattering rates. The $(j+1)$th iteration of $g^{j+1}\left(\varepsilon_{\lambda \textbf{\textit{k}}}\right)$ is considered to satisfy
	\begin{equation}
	g^{j+1}\left(\varepsilon_{\lambda \textbf{\textit{k}}}\right)=\cfrac{S_{\lambda}^{\mathrm{in}}\left(g^{j}\left(\varepsilon_{1 \textbf{\textit{k}}^{\prime}}\right), g^{j}\left(\varepsilon_{-1 \textbf{\textit{k}}^{\prime}}\right)\right)-\cfrac{(-e)E}{\hbar} \cfrac{\partial f_{0}\left(\varepsilon_{\lambda \textbf{\textit{k}}}\right)}{\partial k}}{S_{\lambda}^{\text {out }}+\nu^{\mathrm{el}}},
	\end{equation}
	where we arbitrarily select $g^{0}\left(\varepsilon_{\text {l\textbf{\textit{k}}}}\right)=g^{0}\left(\varepsilon_{-\mathrm{l\textbf{\textit{k}}}}\right)=0.$ In this case, $S_{\lambda}^{\mathrm{in}}=0,$ and Eq.\;(14) provides the first solution as
	\begin{equation}
	g^{1}\left(\varepsilon_{\lambda \textbf{\textit{k}}}\right)=\cfrac{-\cfrac{(-e)E}{\hbar} \cfrac{\partial f_{0}}{\partial k}}{S_{\lambda}^{\text {out }}+\nu^{\mathrm{el}}}.
	\end{equation}
	Equation\;(15) can be regarded as the solution having a relaxation time of $\tau_{\mathrm{f}}=1 /\left(S_{\mathrm{i}}^{\text {out }}+v^{\mathrm{el}}\right)$, which demonstrates that the first step of this iterative process can be regarded as the RTA momentum relaxation rate. However, the iterative process must continue until it converges to an appropriate accuracy. In addition to intrinsic optical phonon modes, remote scattering via surface polar optical phonons (SPOP) modes is known to be a limiting factor of electron mobility in graphene on polar substrates and 2D artificial structures such as silicon metal--oxide--semiconductor field-effect transistors. Although we consider only the intrinsic optical phonons of graphene for the inelastic term, a similar procedure can be applied for the SPOP modes using the quasiparticle scattering rate \cite{Fratini2008a, Chen2008f, Hwang2013e, Lin2014d}. The effect of carrier--carrier scattering on the intraband conductivity appears from the deviation of the carrier distribution from the FD distribution and the asymmetry of the conduction and valence bands with a prominent contribution when the carrier distribution is far from the equilibrium \cite{Ridley2013a}. In this study, we ignore carrier--carrier scattering, because a low electric field causes a small disturbance for the carrier distribution in the 2D-MDF.
	
	\subsection*{B. Iterative solutions of BTE under time-dependent electric field}
	
	Here, we extend Eq.\;(14) regarding the perturbed distribution, $g\left(\varepsilon_{\lambda \textbf{\textit{k}}}\right)$, in the steady state under a constant electric field to include arbitrarily time-dependent driving forces. In this case, $\partial f_{\lambda}(\textbf{\textit{k}}, t) / \partial t \neq 0.$ We explain the derivation of the iterative solution of the BTE by following the procedure in Ref.\,\cite{Willardson1972a}. To include time dependence, a constant $\Omega_{\mathrm{s}} \geq 0$ is added to the scattering-out, and the term, $\Omega_{\mathrm{s}} g^{j}\left(\varepsilon_{\lambda \textbf{\textit{k}}}\right)$, is added to the numerator of Eq.\;(14), which does not affect the solution of $g\left(\varepsilon_{\lambda \textbf{\textit{k}}}\right)$. If there exists a unique $g^{\infty}\left(\varepsilon_{\lambda \textbf{\textit{k}}}\right)$ of
	\begin{equation}
	g^{j+1}\left(\varepsilon_{\lambda \textbf{\textit{k}}}\right)=\cfrac{S_{\lambda}^{\mathrm{in}}-\cfrac{(-e)E}{\hbar} \cfrac{\partial f_{0}}{\partial k}+\Omega_{\mathrm{s}} g^{j}}{S_{\lambda}^{\text {out }}+\nu^{\mathrm{el}}+\Omega_{\mathrm{s}}},
	\end{equation}
	then $g^{\infty}\left(\varepsilon_{\lambda \textbf{\textit{k}}}\right)$ is independent of $\Omega_{\mathrm{s}}$. Furthermore, it is equal to $g^{\infty}\left(\varepsilon_{\lambda \textbf{\textit{k}}}\right)$ for $\Omega_{\mathrm{s}}=0,$ which is the solution to Eq.\;(14). While the condition $\Omega_{\mathrm{s}} \geq 0$ lowers the convergence rate of the sequence, $\left\{g^{j}\left(\varepsilon_{\lambda \textbf{\textit{k}}}\right)\right\}$, because $g^{j+1}\left(\varepsilon_{\lambda \textbf{\textit{k}}}\right)$ approaches $g^{j}\left(\varepsilon_{\lambda \textbf{\textit{k}}}\right)$
	as $\Omega_{\mathrm{s}}$ approaches infinity. Further, we can relate $\displaystyle \lim _{\Omega_{\mathrm{s}} \rightarrow \infty} \Omega_{\mathrm{s}}\left\{g^{j+1}\left(\varepsilon_{\lambda \textbf{\textit{k}}}\right)-g^{j}\left(\varepsilon_{\lambda \textbf{\textit{k}}}\right)\right\}$ to $\partial g_{\lambda}(\varepsilon_{\lambda \textbf{\textit{k}}}) / \partial t$ as follows:
	From Eq.\,(16), 
	\begin{equation}
	\begin{aligned}
	\displaystyle \lim _{\Omega_{\mathrm{s}} \rightarrow \infty} &\Omega_{\mathrm{s}}\left(g^{j+1}\left(\varepsilon_{\lambda \textbf{\textit{k}}}\right)-g^{j}\left(\varepsilon_{\lambda \textbf{\textit{k}}}\right)\right) \\
	=&S_{\lambda}^{\mathrm{in}}-\left(S_{\lambda}^{\mathrm{out}}+\nu^{\mathrm{el}}\right) g^{j+1}\left(\varepsilon_{\lambda \textbf{\textit{k}}}\right)-\frac{(-e)E}{\hbar} \frac{\partial f_{0}}{\partial k} \\
	=&S_{\lambda}^{\mathrm{in}}-\left(S_{\lambda}^{\mathrm{out}}+\nu^{\mathrm{el}}\right) g^{j}\left(\varepsilon_{\lambda \textbf{\textit{k}}}\right)-\frac{(-e)E}{\hbar} \frac{\partial f_{0}}{\partial k},
	\end{aligned}
	\end{equation}
	where the final equation follows from the fact that $g^{j+1}\left(\varepsilon_{\lambda \textbf{\textit{k}}}\right)$ is indistinguishable from $g^{j}\left(\varepsilon_{\lambda \textbf{\textit{k}}}\right)$ when $\Omega_{s}$ approaches infinity. Recalling the definitions of $S_{\lambda}^{\mathrm{in}}$, $S_{\lambda}^{\text {out }}$, and $\nu^{\mathrm{el}}$ in Eqs.\;(11)--(13), we have the Boltzmann equation 
	\begin{equation}\begin{aligned}
	\lim _{\Omega_{\mathrm{s}} \rightarrow \infty} &\Omega_{\mathrm{s}}\left(g^{j+1}\left(\varepsilon_{\lambda \textbf{\textit{k}}}\right) \cos \alpha-g^{j}\left(\varepsilon_{\lambda \textbf{\textit{k}}}\right) \cos \alpha\right)\\
	&=-\frac{(-e)}{\hbar} \textbf{\textit{E}} \frac{\partial}{\partial \textbf{\textit{k}}} f_{\lambda}^{j}(\textbf{\textit{k}})+\sum_{\eta, \lambda^{\prime}} C_{\lambda \lambda^{\prime}}^{\eta}+C_{\lambda}^{\mathrm{el}},
	\end{aligned}\end{equation}
	where $f_{\lambda}^{j}(\textbf{\textit{k}})=f_{0}\left(\varepsilon_{\lambda \textbf{\textit{k}}}\right)+g^{j}\left(\varepsilon_{\lambda \textbf{\textit{k}}}\right) \cos \alpha.$ The left-hand side of Eq.\;(18) is simply identified by $\partial f_{\lambda}^{j}(\textbf{\textit{k}}) / \partial t$ at time $t_{j}=j / \Omega_{\mathrm{s}}$, where $1 / \Omega_{\mathrm{s}}$ is the time increment between successive iterations. Therefore, the sequence, $\{g^{j}\left(\varepsilon_{\lambda \textbf{\textit{k}}}\right)\}$, yields $f_{\lambda}^{j}(\textbf{\textit{k}})$ versus time when $\Omega_{\mathrm{s}}$ is sufficiently large, compared with $S_{\lambda}^{\text {out }}+\nu^{\text {e }}$. Further, Eqs.\;(17) and (18) may be adopted with a slight modification to include the arbitrarily time-dependent electric field, $\textbf{\textit{E}}(t)$. In this instance, $\textbf{\textit{E}}(t)$ becomes a function of time through the iteration index, $j$, such that
	\begin{equation}\begin{aligned}
	g^{j+1}&\left(\varepsilon_{\lambda \textbf{\textit{k}}}\right)=\cfrac{S_{\lambda}^{\mathrm{in}}-\cfrac{(-e)E \left(t_{j}\right)}{\hbar} \cfrac{\partial f_{0}}{\partial k}+\Omega_{\mathrm{s}} g^{j}}{S_{\lambda}^{\mathrm{out}}+\nu^{\mathrm{el}}+\Omega_{\mathrm{s}}}.
	\end{aligned}\end{equation}
	The sequence, $\{g^{j}\left(\varepsilon_{\lambda \textbf{\textit{k}}}\right)\}$, is used to calculate the field-induced current density, $J(t)$, in graphene as
	\begin{equation}
	J\left(t_{j}\right)=\sum_{\lambda} \cfrac{(-e)v_{\mathrm{F}}}{2} \int_{-\infty}^{\infty} N\left(\varepsilon_{\lambda \textbf{\textit{k}}}\right) g^{j}\left(\varepsilon_{\lambda \textbf{\textit{k}}}\right) d \varepsilon_{\lambda \textbf{\textit{k}}},
	\end{equation}
	where $v_{\mathrm{F}}$ is the Fermi velocity, and $N\left(\varepsilon_{\lambda \textbf{\textit{k}}}\right)=2\left|\varepsilon_{\lambda \textbf{\textit{k}}}\right| /\left(\pi \hbar^{2} v_{\mathrm{F}}^{2}\right)$ is the density of states for 2D-MDF. Subsequently, the intraband conductivity can be obtained by
	\begin{equation}
	\sigma(\omega)=\frac{J(\omega)}{E(\omega)},
	\end{equation}
	where $J(\omega)$ and $E(\omega)$ are the Fourier transformations of $J(t)$ and $E(t),$ respectively. Because $g^{j}\left(\varepsilon_{\lambda \textbf{\textit{k}}}\right)$ is a function of electron energy $\varepsilon_{\lambda \textbf{\textit{k}}},$ the computation cost in the proposed method is decreased significantly compared with the full calculation of $f_{\lambda}(\textbf{\textit{k}}, t)$ when solving Eq.\;(1) in a 2D momentum space. The convergence of Eq.\;(14) and (19) is demonstrated by calculating the direct- and alternating-current conductivities in Appendix E. Note that Eqs.\;(14) and (19) are valid under a sufficiently weak electric field, which causes the small distribution changes, $g \left(\varepsilon_{\lambda \textbf{\textit{k}}}\right)$, from the corresponding equilibrium state.
	
	\subsection*{C. Calculation of carrier temperature and optical phonon occupations in hot-carrier cooling process}
	
	Next, we explain the THz-conductivity calculation procedure for hot carriers in photoexcited graphene at quasi-equilibrium. Because the iterative solution of Eq.\;(19) for the BTE is valid under near equilibrium, it cannot be used directly for calculating the hot-carrier distribution at quasi-equilibrium, which requires the thermalized electron and optical phonon distributions in the cooling process following photoexcitation. During the cooling process, 
	hot carriers lose their energy by emitting strongly coupled optical phonons (i.e., $\Gamma$-LO, $\Gamma$-TO, and $\mathbf{K}$) resulting in a change in the optical-phonon occupation. To consider the cooling process for the hot carriers, we employ a comprehensive model based on the rate equations that describe the temporal evolution of the electron temperature, $T_{\mathrm{e}}$, and the optical phonon occupations, $n_{\eta}$ \cite{Wu2012a, Song2012a, Someya2017b, Wang2010a} (for details, see Appendix C).
	\begin{equation}
	\frac{d T_{\mathrm{e}}}{d t}=\frac{I_{\mathrm{p}}}{C}-\frac{\sum_{\eta} R_{\eta}^{\mathrm{Net}} \hbar \omega_{\eta}}{C}-\frac{J_{\mathrm{sc}}}{C}\\,
	\end{equation}
	\begin{equation}
	\frac{d n_{\Gamma\mathchar`-\mathrm{LO}}}{d t}=R_{\mathrm{M}, \Gamma\mathchar`-\mathrm{LO}}^{\mathrm{Net}}-\frac{n_{\Gamma\mathchar`-\mathrm{LO}}-n_{\Gamma0}}{\tau_{\mathrm{ph}}},
	\end{equation}
	\begin{equation}
	\frac{d n_{\Gamma\mathchar`-\mathrm{TO}}}{d t}=R_{\mathrm{M}, \Gamma\mathchar`-\mathrm{TO}}^{\mathrm{Net}}-\frac{n_{\Gamma\mathchar`-\mathrm{TO}}-n_{\Gamma0}}{\tau_{\mathrm{ph}}},
	\end{equation}
	\begin{equation}
	\frac{d n_{\mathrm{\textbf{K}}}}{d t}=R_{\mathrm{M}, \mathrm{\textbf{K}}}^{\mathrm{Net}}-\frac{n_{\mathrm{\textbf{K}}}-n_{\mathrm{\textbf{K}}0}}{\tau_{\mathrm{ph}}}.
	\end{equation}
	In this case, $I_{\mathrm{p}}$ represents the energy injected into the graphene sample during laser irradiation, which is assumed to be of the hyperbolic secant form, $I_{\mathrm{p}}(t)=\left(F_{\mathrm{ab}} / 2 \tau_{\mathrm{exc}}\right) \operatorname{sech}^{2}\left(t / \tau_{\mathrm{exc}}\right)$, where $F_{\mathrm{ab}}$ is the absorbed pump fluence, and $2 \tau_{\operatorname{exc}}$ is the pump-pulse duration. Furthermore, $C$ is the sum of the specific heat of the electrons in the conduction and valence bands. $R_{\eta}^{\mathrm{Net}}=R_{\eta}-G_{\eta}$ denotes the total balance between the optical phonon emission and absorption rate, and $J_{\mathrm{sc}}$ denotes the energy-loss rate for the SC carrier-cooling process \cite{Song2012a}. $R_{M, \eta}^{\text {Net }}=R_{\mathrm{M}, \eta}-G_{\mathrm{M}, \eta}$ denotes the total balance between the optical phonon emission and absorption rate per number of phonon modes. Moreover, $n_{\Gamma0}$ and $n_{\mathrm{\textbf{K}}0}$ represent the phonon occupation near $\Gamma$ and \textbf{K} points, respectively, in equilibrium at room temperature, and $\tau_{\mathrm{ph}}$ is the phenomenological phonon decay time via the phonon--phonon interaction caused by lattice anharmonicity. The acoustic phonon occupation is assumed to remain unchanged from the equilibrium state for the picosecond time range following photoexcitation \cite{Johannsen2013b}. By substituting the solution of $T_{\mathrm{e}}$ and $n_{\eta}$ obtained by numerically solving the coupled Eqs.\;(22)--(25) into Eq.\;(19) during the iteration process, the sequence, $\left\{g^{j}\left(\varepsilon_{\lambda \textbf{\textit{k}}}\right)\right\}$, for the disturbed distribution caused by the applied THz electric field, $E(t)$, which includes the temporal evolution of $T_{\mathrm{e}}$ and $n_{\eta}$ in the cooling process can be obtained. The solution of Eq.\;(19) using $T_{\mathrm{e}}$ and $n_{\eta}$ is not valid for the THz-conductivity calculation of hot carriers with a highly nonequilibrium distribution (non-FD distribution) or hot FD distributions with separate quasi-Fermi levels, because Eq.\;(3) assumes $f_{0}\left(\varepsilon_{\lambda \textbf{\textit{k}}}\right)$ to be the equilibrium or quasi-equilibrium FD distribution with a single chemical potential.
	
	\section{Numerical results}
	In this section, we present the numerical simulations performed to investigate the intrinsic and extrinsic carrier scatterings on the intraband optical conductivity of hot carriers in monolayer graphene having different carrier concentrations. We considered intrinsic carrier-scattering mechanisms caused by intrinsic optical and acoustic phonons and the extrinsic scattering mechanisms caused by the charged impurities on the substrate and weak scatterers, such as defects and neutral impurities \cite{Lin2014d}. 
	\begin{figure}[b]
		\centering
		\includegraphics[width=8.6cm,bb=0 0 283 191]{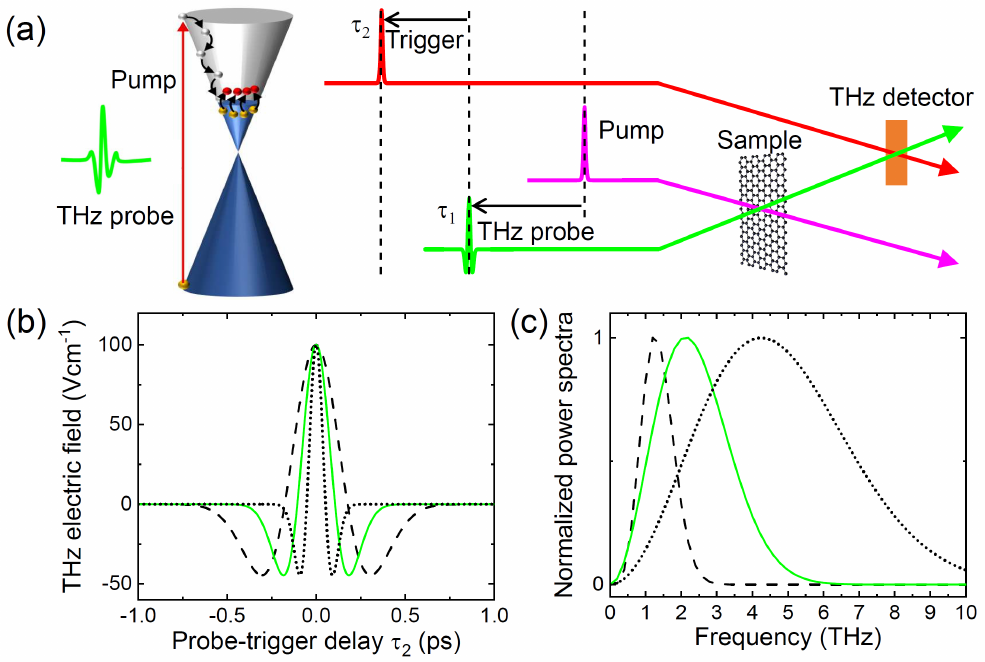}
		\caption{\label{fig1} (a) Schematic representation of OPTP experiment used in the simulation. The sample was probed at different times following photoexcitation by varying $\tau_{\mathrm{1}}$. Varying $\tau_{\mathrm{2}}$ enabled the electric field, $\textbf{\textit{E}}_{\mathrm{t}}(\tau_{\mathrm{1}},\tau_{\mathrm{2}}$), of the THz-probe pulse to be sampled by the trigger pulse. (b) The temporal waveforms of the THz-probe pulse with pulse durations of $2\tau_{p} = 150$ (black dotted line), 300 (green solid line) and $500\,\mathrm{fs}$ (black broken line). (c) The corresponding normalized FFT spectra of the THz-probe pulse.}
	\end{figure}
	
	Figure\;1(a) presents a schematic of the OPTP experimental setup used in the simulation. An optical pump pulse having a temporal duration of $2 \tau_{\mathrm{exc}}=35\,\mathrm{fs}$ irradiates a graphene sample at room temperature ($ T_0= 295\,\mathrm{K}$) on a substrate at a normal incidence angle. The dielectric constants of the substrate are listed in Table I. Owing to the presence of the substrate, $1.37\%$ of the incident pump pulse is absorbed in the graphene layer if we neglect the saturable absorption of pump pulses in graphene \cite{Kumar2009a, Marini2017a}. The created nonequilibrium electron and hole pairs are rapidly recombined by the Auger recombination process and interband optical phonon scattering. The nonequilibrium distribution is assumed to change into the quasi-equilibrium hot-carrier state with a single chemical potential, at which point the present procedure can be applied. In doped graphene, the rapid recombination of photoexcited carriers within $130\,\mathrm{fs}$ is caused by the limited phase space of the impact ionization process, as reported in Ref \cite{Gierz2013c}. Figure\;1(b) presents the temporal waveforms of THz-probe pulse calculated by the second derivative of the Gaussian function, $\exp (-t^{2} / \tau_{p}^{2})$, with pulse durations of $2\tau_{p}=150, 300$ and $500\,\mathrm{fs}$. Figure.\;1(c) presents the corresponding normalized FFT power spectra that have center frequencies $\omega / 2 \pi$ of 4.2, 2.2, and $1.2\,\mathrm{THz}$. The THz-probe pulses are transmitted to the graphene sample at a normal incidence angle at time $\tau_{1}$ after photoexcitation, and the waveforms of the transmitted THz pulses are assumed to be detected using a time-resolved detection scheme, such as via electro-optic sampling, by varying the delay time, $\tau_{2}$, between the THz probe and the trigger pulses. The parameters of the graphene sample and experimental setups used in the simulation are summarized in Table\;I and Figs.\;A.1--C.3 in Appendices\;A--C.
	\begin{table}[b]
		\caption{\label{tab:fonts} Parameters of graphene with different carrier concentrations and experimental setups used for simulation.}
		\begin{ruledtabular}
			\squeezetable
			\begin{tabular}{lc}
				Quantity & Values \\
				\hline Fermi energy $\varepsilon_{\mathrm{F}}\,(\mathrm{eV})$ & $-0.01, -0.15, -0.43$ \\
				Fermi velocity $v_{\mathrm{F}}\,(\mathrm{ms}^{-1})$ & $1.1 \times 10^{6}$ \\
				Static dielectric constant & \multirow{2}{*}{3.0} \\
				of substrate $\epsilon_{\mathrm{s}}$\\
				Dielectric constant of substrate & \multirow{2}{*}{3.0} \\
				at THz-probe wavelength $\epsilon_{\mathrm{THz}}$ \\
				Charged impurity concentration & \multirow{2}{*}{0.1, 1.0} \\
				$n_{\mathrm{i}}(10^{12}\,\mathrm{cm}^{-2})$ \\
				Resistivity of weak scatterers $\rho_{\mathrm{s}}\,(\Omega)$&100\\
				Deformational potential of&\multirow{2}{*}{30.0}\\
				acoustic phonon $D_{\mathrm{ac}}\,(\mathrm{eV})$\\
				EPC coefficient at $\Gamma$ point&\multirow{2}{*}{45.6}\\
				$\left\langle D_{\mathrm{\Gamma}}^{2}\right\rangle_{\mathrm{F}}\,(\mathrm{eV} \mathrm{\AA}^{-1})^{2}$\\
				EPC coefficient at \textbf{K} point&\multirow{2}{*}{92.1}\\
				$\left\langle D_{\mathrm{\textbf{K}}}^{2}\right\rangle_{\mathrm{F}}\,(\mathrm{eV} \mathrm{\AA}^{-1})^{2}$\\
				Optical phonon decay time $\tau_{\mathrm{ph}}\,(\mathrm{ps})$&1.0\\
				Pulse duration of pump pulse $2\tau_{\mathrm{exc}}\,(\mathrm{fs})$&35\\
				Pulse duration of THz probe $2\tau_{\mathrm{p}}\,(\mathrm{fs})$&150, 300, 500\\
			\end{tabular}
		\end{ruledtabular}
	\end{table}
	
	\subsection*{A. Undoped graphene}
	First, we present the numerical results of undoped graphene with different charged impurity concentrations of $n_{\mathrm{iL}}=0.1$ and $n_{\mathrm{iH}}=1.0 \times 10^{12}$ $\mathrm{cm}^{-2}$, as illustrated in Figs.\;2 and 3, respectively. The charge inhomogeneity and disorder in graphene smear out the intrinsic behavior near the Dirac point, thereby resulting in unintended carrier doping and a finite $\varepsilon_{\mathrm{F}}$ \cite{Adam2007b}. Thus, we set the finite $p$-type carrier concentration as $\varepsilon_{\mathrm{F}}=-0.01\,\mathrm{eV}$ for the undoped graphene \cite{Kar2014a}. The corresponding hole concentration at $T=0\,\mathrm{K}$ is $n_{\mathrm{c}}=7.0 \times 10^{9}$ $\mathrm{cm}^{-2}$. The calculated DC conductivity of the undoped graphene with $n_{\mathrm{iL}}$ and $n_{\mathrm{iH}}$ at equilibrium without pump fluence are approximately $\sigma_{\mathrm{DC}}=6G_{0}$ and $0.8G_{0}$, respectively, where $G_{0}=2 e^{2} / h$ is the quantum conductance. 
	\begin{figure}[b]
		\centering
		\includegraphics[width=5.5cm, bb=0 0 283 406]{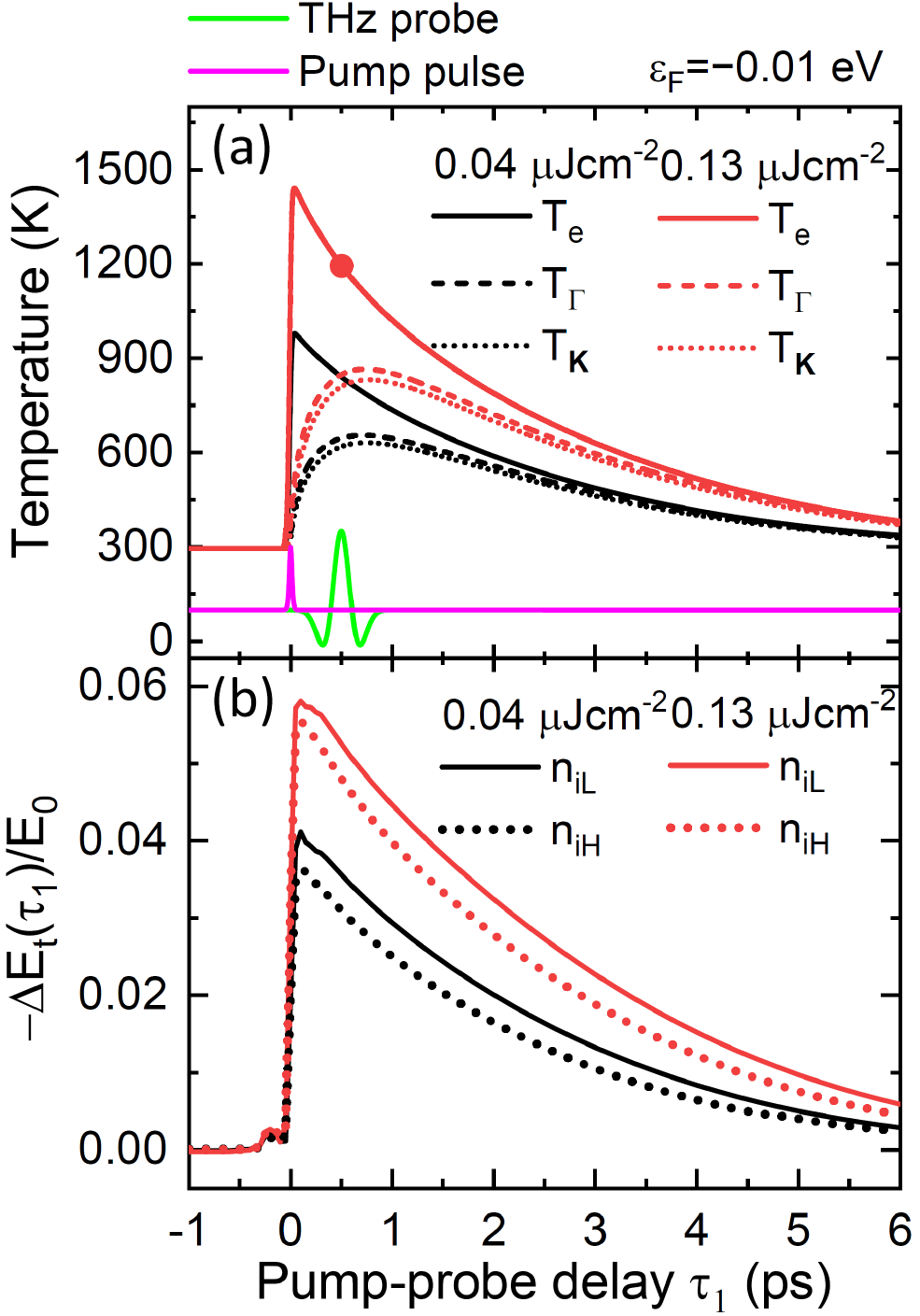}
		\caption{\label{fig2} (a) Temporal evolutions of $T_{\mathrm{e}}$ and $T_{\eta}$ of undoped graphene for $F_{\mathrm{ab}}=0.04$ and $0.13\,\mathrm{\upmu Jcm}^{-2}$. The optical phonon temperatures were calculated by inverting the Bose--Einstein distribution function, $n_{\eta}=1 /(e^{\hbar \omega_{\eta} / k_{\mathrm{B}} T_{\eta}}-1)$. (b) Temporal evolutions of $-\Delta E_{\mathrm{t}}\left(\tau_{1}\right) / E_{0}$ calculated using the THz-probe pulse with $2 \tau_{\mathrm{p}}=300\,\mathrm{fs}$ for $n_{\mathrm{iL}}$ and $n_{\mathrm{iH}}$.}
	\end{figure}
	
	Figure\;2(a) presents the temporal evolutions of the carrier temperature ($T_{\mathrm{e}}$) and the optical phonon temperatures ($T_{\Gamma}, T_{\textbf{K}}$) in the undoped graphene under absorbed pump fluences of $F_{\mathrm{ab}}=0.04$ and $0.13\,\mathrm{\upmu Jcm}^{-2}$. 
	Here, the charged impurity scattering is elastic and does not affect the temporal evolutions of $T_{\mathrm{e}}$ and $T_{\mathrm{\eta}}$.
	After the photoexcitation, $T_{\mathrm{e}}$ increased to almost 1,000 and $1,500\,\mathrm{K}$ and relaxed to the equilibrium via double exponential decay with fast decay times of $\tau_{\mathrm{T} 1}=0.25$ and $0.37\,\mathrm{ps}$ and slow decay times of $\tau_{\mathrm{T} 2}=2.6$ and $2.8\,\mathrm{ps}$ for $F_{\mathrm{ab}}=0.04$ and $0.13\,\upmu \mathrm{J} \mathrm{cm}^{-2}$, respectively. 
	The fast decay with $\tau_{\mathrm{T} 1}$ corresponds to the hot-carrier relaxation by the increased optical phonon emission process where hot-carrier temperature is larger than those of optical phonon modes. It is mainly governed by the total balance of the energy exchange rate for optical phonon emission and absorption, $R_{\eta}^{\mathrm{Net}}\hbar \omega_{\eta}$ (see Appendix\;C, Fig.\;C.2). The slow decay with $\tau_{\mathrm{T} 2}$ is related to hot-carrier and -phonon relaxation to the equilibrium and depend on the optical phonon decay time, $\tau_{\mathrm{ph}}$, and the energy-loss rate, $J_{\mathrm{sc}}$, via the SC carrier cooling process (see Appendix C, Fig.\;C.1(b)). 
	Figure\;3(a) and (b) show the complex conductivity of undoped graphene, $\sigma\left(\omega, \tau_{1}\right)=\sigma_{1}\left(\omega, \tau_{1}\right)+i \sigma_{2}\left(\omega, \tau_{1}\right)$, with  $n_{\mathrm{iL}}$ and  $n_{\mathrm{iH}}$ calculated using the THz-probe pulse with $2 \tau_{\mathrm{p}}=300\,\mathrm{fs}$. The blue and red symbols are $\sigma\left(\omega, \tau_{1}\right)$ without and with pump fluence, respectively, calculated by the iterative method using the temporal evolutions of the $T_{\mathrm{e}}$, $T_{\Gamma}$, and $T_{\textbf{K}}$ illustrated in Fig.\;\ref{fig2}(a). Both of $\sigma\left(\omega, \tau_{1}\right)$ with $n_{\mathrm{iL}}$ and  $n_{\mathrm{iH}}$ substantially increase by the photoexcitation. The frequency dependence of $\sigma\left(\omega, \tau_{1}\right)$ with  $n_{\mathrm{iL}}$ and  $n_{\mathrm{iH}}$ at equilibrium state show the simple Drude type behavior. However, those at the hot-carrier state show non-Drude behavior, which is remarkable for  $\sigma\left(\omega, \tau_{1}\right)$ with $n_{\mathrm{iH}}$.
	\begin{figure}
		\centering
		\includegraphics[width=8.6cm, bb=0 0 283 253]{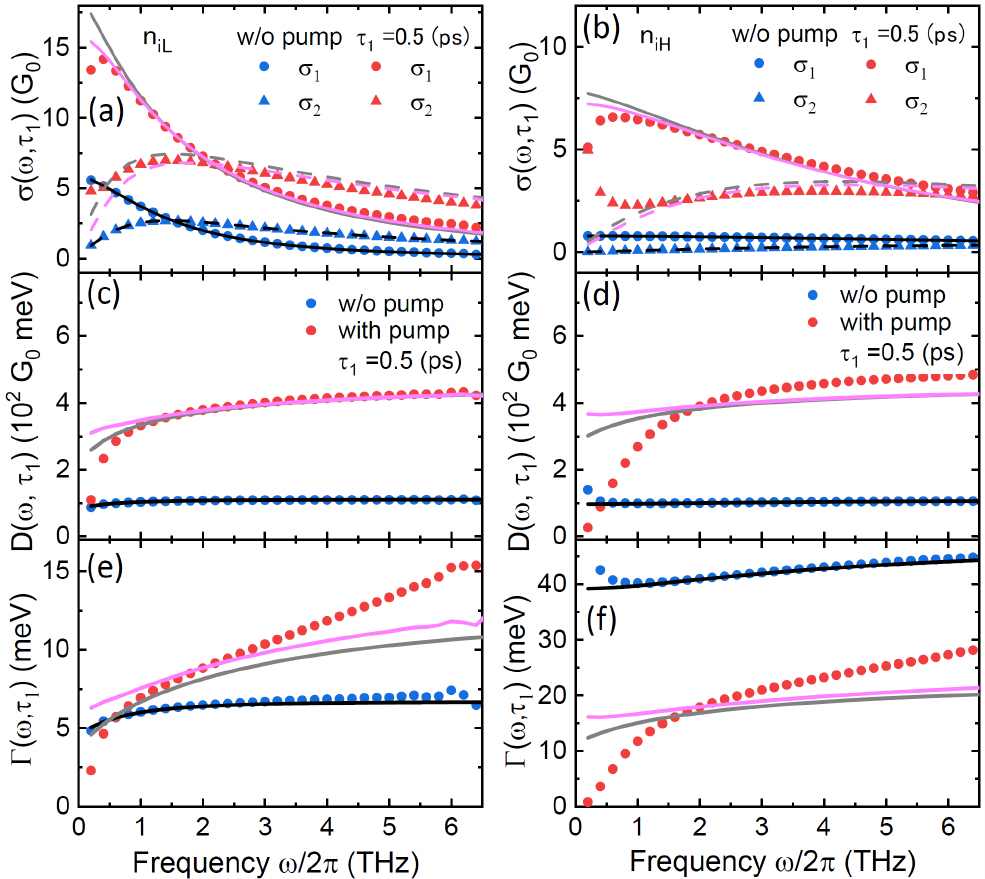}
		\caption{\label{fig3} Simulation results of undoped graphene calculated using the THz-probe pulse with $2 \tau_{\mathrm{p}}=300\,\mathrm{fs}$. $\sigma\left(\omega, \tau_{1}\right)$ for (a) $n_{\mathrm{iL}}$ and (b) $n_{\mathrm{iH}}$, respectively, with $F_{\mathrm{ab}}=0.13\,\upmu \mathrm{J} \mathrm{cm}^{-2}$ at $\tau_{1}=0.5\,\mathrm{ps}$, indicated by the filled solid circle in Fig. 2(a). Frequency dependence of $D\left(\omega, \tau_{1}\right)$ for (c) $n_{\mathrm{iL}}$ and (d)  $n_{\mathrm{iH}}$ and $\Gamma\left(\omega, \tau_{1}\right)$ for (e) $n_{\mathrm{iL}}$ and (f)  $n_{\mathrm{iH}}$. The blue and red symbols are calculated by the iterative method and the pink lines by the iterative method with the fixed temperatures at $\tau_{2}=0\,\mathrm{ps}$. The black and gray lines reflect the RTA formula of Eq. (B10)}
	\end{figure}
		
	To discuss the frequency dependence of $\sigma\left(\omega, \tau_{1}\right)$ and the contribution of the free carrier concentration and scattering rate, we exploited the extended Drude model with the frequency dependent Drude weight, $D\left(\omega, \tau_{1}\right)$, and momentum relaxation rate, $\Gamma\left(\omega, \tau_{1}\right)$, expressed as 
	\begin{equation}\begin{aligned} 
	\sigma\left(\omega, \tau_{1}\right) &=\sigma_{1}\left(\omega, \tau_{1}\right)+i \sigma_{2}\left(\omega, \tau_{1}\right) \\
	&=\frac{1}{\pi} \frac{D\left(\omega, \tau_{1}\right)}{\Gamma\left(\omega, \tau_{1}\right)-i \omega}.
	\end{aligned}\end{equation}
	$D\left(\omega, \tau_{1}\right)$ and $\Gamma\left(\omega, \tau_{1}\right)$ are derived from $\sigma\left(\omega, \tau_{1}\right)$, as follows:
	\begin{equation}
	D\left(\omega, \tau_{1}\right)=\frac{-\pi \omega}{\operatorname{Im}\left[\sigma^{-1}\left(\omega, \tau_{1}\right)\right]},
	\end{equation}
	\begin{equation}
	\Gamma\left(\omega, \tau_{1}\right)=\frac{\sigma_{1}\left(\omega, \tau_{1}\right)}{\sigma_{2}\left(\omega, \tau_{1}\right)} \omega.
	\end{equation}
	Note that a simple Drude model expressed by $\sigma_{\mathrm{D}}(\omega)=D / \pi(\Gamma-i \omega)$ has constant values of the Drude weight, $D$, which is proportional to the 
	free-carrier concentration, and the relaxation rate, $\Gamma$, in the frequency space. Both the frequency dependence of $D\left(\omega, \tau_{1}\right)$ and $\Gamma\left(\omega, \tau_{1}\right)$ in the extended Drude model indicate the deviation of $\sigma\left(\omega, \tau_{1}\right)$ from the simple Drude model. 
		\begin{figure}[t]
		\centering
		\includegraphics[width=8.5cm, bb=0 0 283 193]{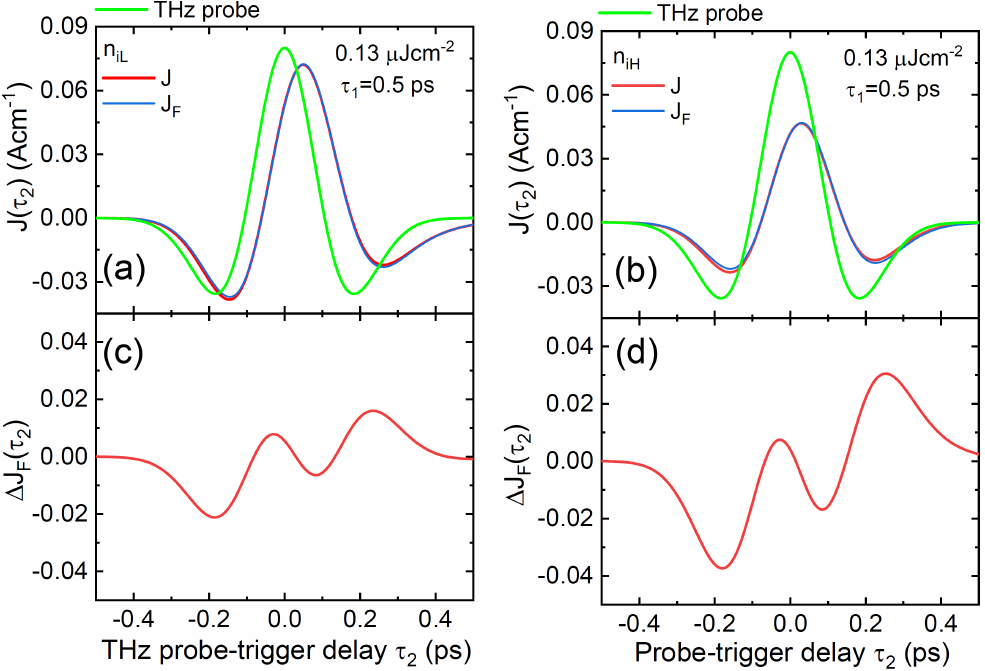}
		\caption{\label{fig4} Temporal waveforms of the THz-probe pulse with $2 \tau_{\mathrm{p}}=300\,\mathrm{fs}$ (green line) and THz field-induced current densities, $J\left(\tau_{2}\right)$ and $J_{\mathrm{F}}\left(\tau_{2}\right)$, in undoped graphene with (a) $n_{\mathrm{iL}}$ and (b) $n_{\mathrm{iH}}$ at $\tau_{1}=0.5\,\mathrm{ps}$ for $F_{\mathrm{ab}}=0.13\,\upmu \mathrm{J} \mathrm{cm}^{-2}$. The red and blue lines indicate $J\left(\tau_{2}\right)$ and $J_{\mathrm{F}}\left(\tau_{2}\right)$ calculated by varied and fixed temperatures with $\tau_{2}$, respectively. The normalized difference in current waveforms, $\Delta J_{\mathrm{F}}\left(\tau_{2}\right)$, for (c) $n_{\mathrm{iL}}$ and 
		(d) $n_{\mathrm{iH}}$, where $\Delta J_{\mathrm{F}}\left(\tau_{2}\right)=(J\left(\tau_{2}\right)-J_{\mathrm{F}}) / J_{\max }$.}
	\end{figure}
	
	Figures\;3(c)--(f) present the frequency dependence of $D\left(\omega, \tau_{1}\right)$ and  $\Gamma\left(\omega, \tau_{1}\right)$ with $n_{\mathrm{iL}}$ and $n_{\mathrm{iH}}$. 
	For both cases, the $D\left(\omega, \tau_{1}\right)$ at $\tau=0.5\,\mathrm{ps}$ increases to four times equilibrium, indicating a substantial increase of free-carrier concentration.
	On the contrary, $\Gamma\left(\omega, \tau_{1}\right)$ with $n_{\mathrm{iL}}$ and $n_{\mathrm{iH}}$ show different behaviors. 
	$\Gamma\left(\omega, \tau_{1}\right)$ with $n_{\mathrm{iL}}$ increases to approximately two times equilibrium, indicating the increased optical phonon scattering (see Fig.\;B.1 in Appendix B), and $\Gamma\left(\omega, \tau_{1}\right)$ with $n_{\mathrm{iH}}$ decreases to half of that at equilibrium, owing to the reduced charged impurity scattering, which is caused by the carrier-screening effect, decreasing with increasing free-carrier concentration.
	As a result, the conductivity enhancement by the photoexcitation is more remarkable for $n_{\mathrm{iH}}$, as seen in Figs. 3(a) and (b). 
	
	The frequency dependence of $\sigma\left(\omega, \tau_{1}\right)$ in the hot-carrier state strongly depends on the charged impurity concentration. For $n_{\mathrm{iH}}$, the non-Drude behavior is remarkable, where the $\sigma_{1}\left(\omega, \tau_{1}\right)$/$\sigma_{2}\left(\omega, \tau_{1}\right)$ decreases/increases with decreasing $\omega$ below $\omega/2 \pi=1\,\mathrm{THz}$ and might be attributed to the large temporal variation of the charged impurity scattering, because the THz probe pulse is broad and insufficient to capture the hot-carrier state at $\tau_{1}=0.5\,\mathrm{ps}$ instantaneously. To verify the effect, we show THz-field-induced current densities, $J$ and $J_{\mathrm{F}}$, calculated using the iterative method with varied and fixed the temperatures with $\tau_2$, respectively, in Fig. 4. The deformation and phase shift of $J$ from the THz pulse waveform reflects the temporal evolution of the hot-carrier distribution and scattering rate. It is seen in Fig. 4(c) and (d) that the normalized difference, $\Delta J_{\mathrm{F}}=(J-J_{\mathrm{F}})/J_{\mathrm{max}}$, for $n_{\mathrm{iH}}$ is twice as large as that for $n_{\mathrm{iL}}$ owing to the strong charged impurity scattering, which quickly changes, depending on the carrier temperature via carrier screening effect. Because the $\sigma_{\mathrm{F}}\left(\omega, \tau_{1}\right)=\sigma_{\mathrm{F1}}\left(\omega, \tau_{1}\right)+i \sigma_{\mathrm{F2}}\left(\omega, \tau_{1}\right)$, calculated with the fixed temperatures (pink lines in Fig. 3(a) and (b)), show the behaviors close to the simple Drude model, the non-Drude behaviors are mainly attributed to the temporal variations of the hot-carrier distribution and scattering rate. For comparison, we also plot the $\sigma_{\mathrm{RTA}}\left(\omega, \tau_{1}\right)$ (black and gray lines) obtained by the RTA calculation (see Appendix B). Although the RTA calculations well-agree with the iterative method in the equilibrium state, it fails to reproduce the hot-carrier $\sigma\left(\omega, \tau_{1}\right)$ with both fixed and varied temperatures. The deviation of the RTA calculation from the $\sigma\left(\omega, \tau_{1}\right)$ is obvious and remarkable in the low frequency where the THz field probes the transient hot-carrier state for a longer duration than that in a high frequency. 
	Note that the difference between the RTA calculation and the iterative method having fixed temperature is simply caused by the difference of the energy dependence of the carrier relaxation rate. 
	
	To present the comparison of temporal evolutions between $T_{\mathrm{e}}$ and the transmission change, $-\Delta E_{\mathrm{t}}\left(\tau_{1}\right) / E_{0}$, defined as $-\left(E_{\mathrm{t}}\left(\tau_{2}, \tau_{1}\right)-E_{t}\left(\tau_{2}\right)\right) / E_{t}\left(\tau_{2}\right)$ at $\tau_{2}=0\,\mathrm{ps}$ when the electric field of the THz-probe pulse exhibits the maximum amplitude, we show in Fig. 2(b) the results of  our theoretical calculations (for details, see Appendix D). 
	Here, $E_{t}\left(\tau_{2}\right)$ is the THz electric field transmitted through the graphene without photoexcitation, and $-\Delta E_{\mathrm{t}}\left(\tau_{1}\right) / E_{0}$ was calculated using the THz-probe pulse with $2\tau_{\mathrm{p}}=300\,\mathrm{fs}$. 
	The $-\Delta E_{\mathrm{t}}\left(\tau_{1}\right) / E_{0}$ is useful for discussing the hot-carrier relaxation and photoconductivity, $\Delta \sigma\left(\omega, \tau_{1}\right)=\sigma\left(\omega, \tau_{1}\right)-\sigma_{0}(\omega)=\Delta \sigma_{1}\left(\omega, \tau_{1}\right)+ i \Delta \sigma_{2}\left(\omega, \tau_{1}\right)$, around the center frequency of the THz-probe pulse, where $\sigma_{0}(\omega)$ is the intra-band optical conductivity of graphene without pump fluence. 
	$-\Delta E_{\mathrm{t}}\left(\tau_{1}\right) / E_{0}>0$ and $-\Delta E_{\mathrm{t}}\left(\tau_{1}\right) / E_{0}<0$ indicate positive and negative photoconductivities, $\Delta \sigma_{1}\left(\omega, \tau_{1}\right)$, respectively. 
	As can be observed in Fig.\;2(b), the positive photoconductivity of the undoped graphene appear for both $F_{\mathrm{ab}}=0.04$ and $0.13\,\upmu \mathrm{Jcm}^{-2}$, as reported in Ref.\,\cite{Frenzel2014d, Kar2014a}. However, the change of magnitudes and their decay times depend on $n_{\mathrm{i}}$. For $n_{\mathrm{iL}}$, the relaxation curves show nearly single exponential decays. The decay times are $\tau_{\mathrm{THz}}=2.4$ and $3.1\,\mathrm{ps}$ for $F_{\mathrm{ab}}=0.04$ and $0.13\,\upmu \mathrm{Jcm}^{-2}$, respectively. For $n_{\mathrm{iH}}$, the $-\Delta E_{\mathrm{t}}\left(\tau_{1}\right) / E_{0}$ curves also exhibit nearly single exponential decays with $\tau_{\mathrm{THz}}=2.9$ and $3.7\,\mathrm{ps}$ for $F_{\mathrm{ab}}=0.04$ and $0.13\,\upmu \mathrm{Jcm}^{-2}$, respectively. These results indicate that the slow decay times, $\tau_{\mathrm{T2}}$, of hot carriers can be approximately estimated from $\tau_{\mathrm{THz}}$ in the case of undoped graphene. The disappearance of the fast decay is caused by the cancellation of the positive and negative contributions of temperature-dependent Drude weight and carrier-scattering rate for $\sigma\left(\omega, \tau_{1}\right)$ during the initial stage after photoexcitation.

	\subsection*{B. Heavily doped graphene}
	Next, we consider the heavily \textit{p}-type doped graphene with $\varepsilon_{\mathrm{F}}=-0.43\,\mathrm{eV}$ (the corresponding hole concentration at $T=0\,\mathrm{K}$ is $n_{\mathrm{c}}=1.3 \times 10^{13}\,\mathrm{cm}^{-2}$) having charged impurity concentrations set to $n_{\mathrm{iL}}=0.1 \times 10^{12}\,\mathrm{cm}^{-2}$ and $n_{\mathrm{iH}}=1.0 \times 10^{12}\,\mathrm{cm}^{-2}$. 
	The corresponding DC conductivities at $T_{0}=295\,K$ are $\sigma_{\mathrm{DC}}=$ $21.5G_{0}$ and $18.0G_{0}$, respectively. The numerical results are illustrated in Figs.\;5 and 6. 
	\begin{figure}[b]
		\centering
		\includegraphics[width=5.5cm, bb=0 0 283 399]{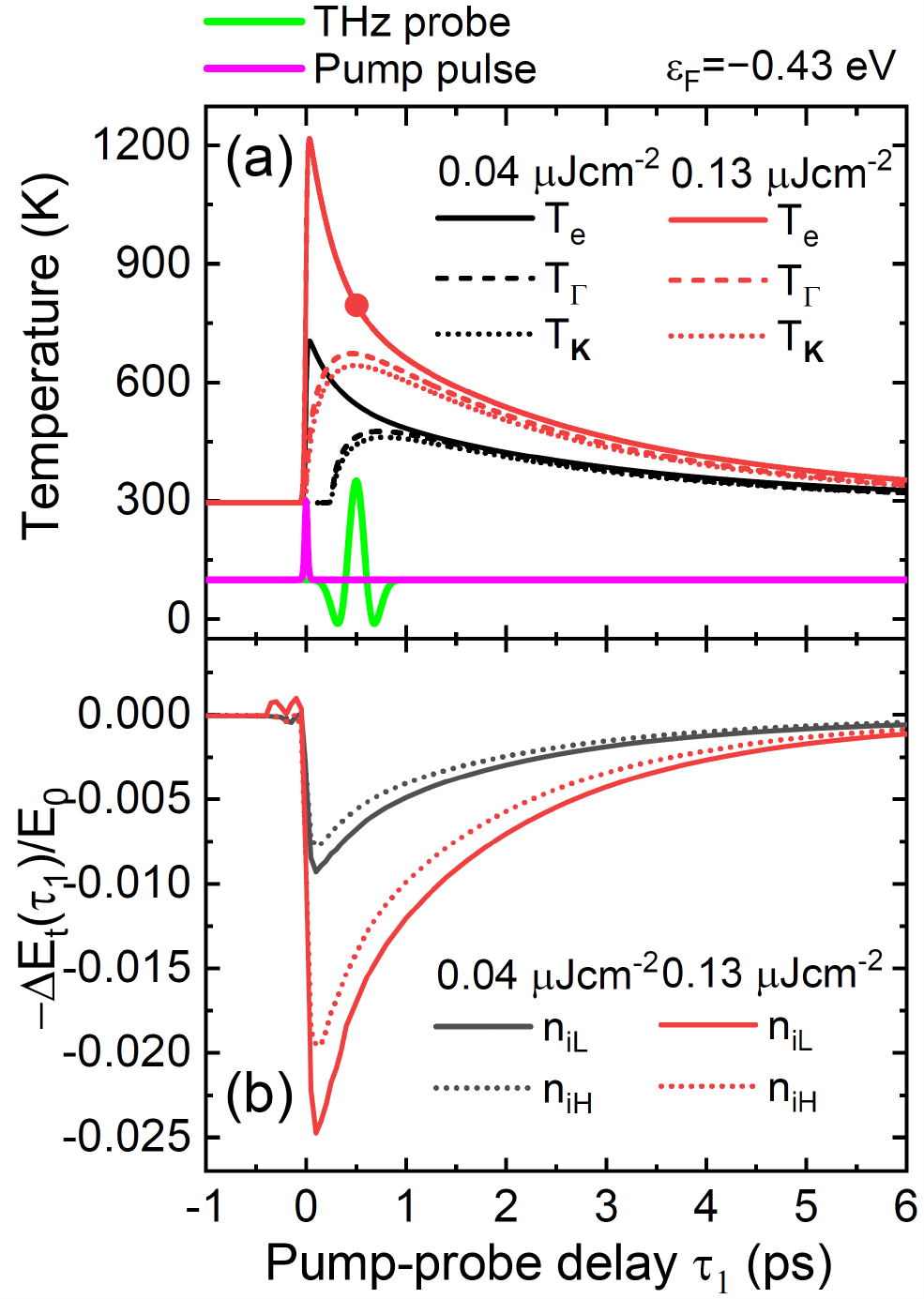}
		\caption{\label{fig6}  Temporal evolutions of $T_{\mathrm{e}}$ and $T_{\eta}$ of heavily doped graphene for $F_{\mathrm{ab}}=0.04$ and $0.13\,\mathrm{\upmu Jcm}^{-2}$. (b) Temporal evolutions of $-\Delta E_{\mathrm{t}}\left(\tau_{1}\right) / E_{0}$ calculated using the THz-probe pulse with $2 \tau_{\mathrm{p}}=300\,\mathrm{fs}$ for $n_{\mathrm{iL}}$ and $n_{\mathrm{iH}}$.}
	\end{figure}
	\begin{figure}[t]
		\centering
		\includegraphics[width=8.6cm, bb=0 0 283 250]{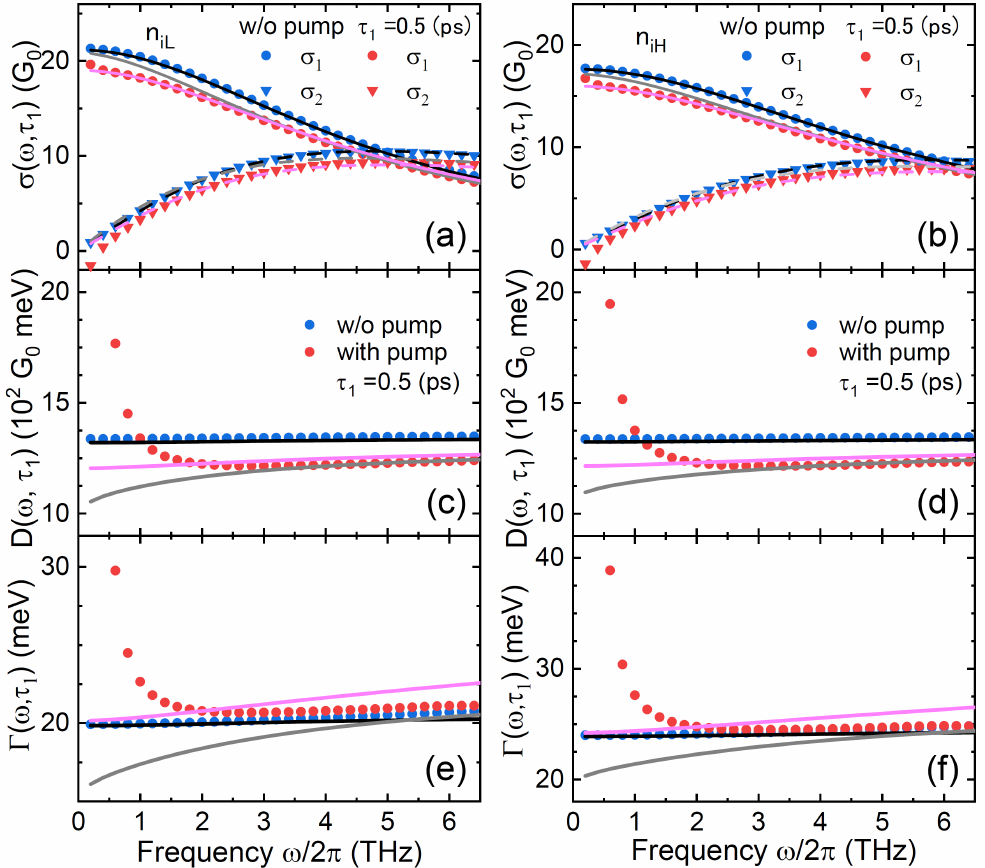}
		\caption{\label{fig7} Simulation results of heavily doped graphene calculated using the THz-probe pulse with $2 \tau_{\mathrm{p}}=300\,\mathrm{fs}$. $\sigma\left(\omega, \tau_{1}\right)$ for (a)  $n_{\mathrm{iL}}$ and (n) $n_{\mathrm{iH}}$ , respectively with $F_{\mathrm{ab}}=0.13\,\upmu \mathrm{J} \mathrm{cm}^{-2}$ at $\tau_{1}=0.5\,\mathrm{ps}$, indicated by the filled solid circle in Fig. 5(a). Frequency dependence of $D\left(\omega, \tau_{1}\right)$ for (c) $n_{\mathrm{iL}}$ and (d)  $n_{\mathrm{iH}}$.  Frequency dependence of $\Gamma\left(\omega, \tau_{1}\right)$ for (e)) $n_{\mathrm{iL}}$ and (f)  $n_{\mathrm{iH}}$. The blue and red symbols are calculated using the iterative method with the varied temperatures and the pink lines by the iterative method with the fixed temperatures at $\tau_{2}=0\,\mathrm{ps}$. The black and gray lines reflect the RTA formula of Eq. (B10).}
	\end{figure}
	\begin{figure}[t]
		\centering
		\includegraphics[width=8.6cm, bb=0 0 283 189]{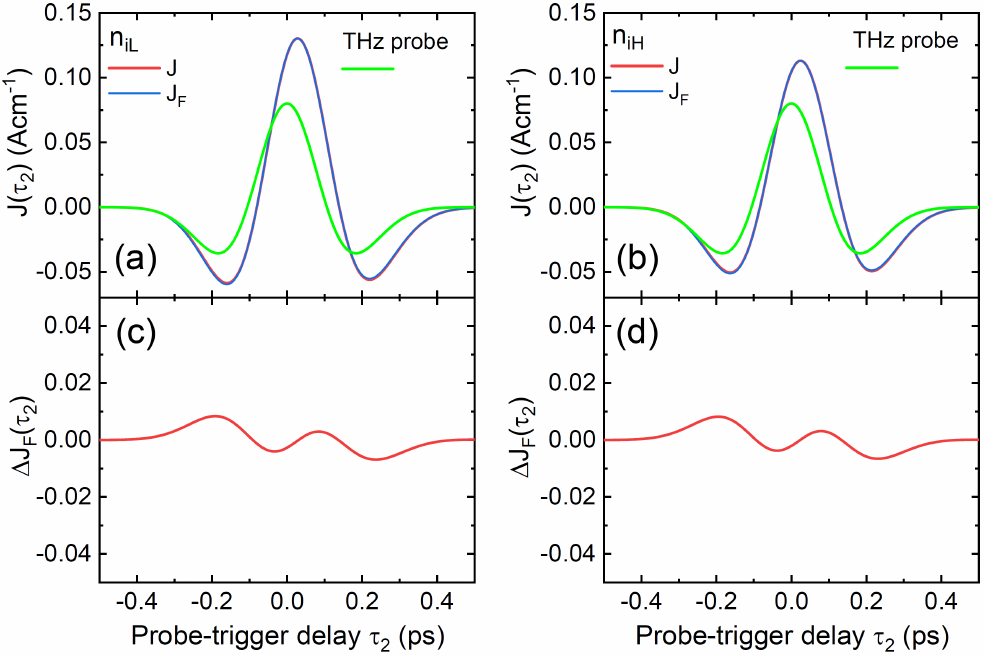}
		\caption{\label{fig8} Temporal waveforms of the THz-probe pulse having $2 \tau_{\mathrm{p}}=300\,\mathrm{fs}$ (green line) and THz field-induced current densities, $J\left(\tau_{2}\right)$ and $J_{\mathrm{F}}\left(\tau_{2}\right)$, in heavily doped graphene with (a) $n_{\mathrm{iL}}=0.1 \times 10^{12}\,\mathrm{cm}^{-2}$ and (b) $n_{\mathrm{iH}}=1.0 \times 10^{12}\,\mathrm{cm}^{-2}$ at $\tau_{1}=0.5\,\mathrm{ps}$ for $F_{\mathrm{ab}}=0.13\,\upmu \mathrm{J} \mathrm{cm}^{-2}$. The red and blue lines indicate $J\left(\tau_{2}\right)$ and $J_{\mathrm{F}}\left(\tau_{2}\right)$, respectively. The normalized difference in current waveforms $\Delta J_{\mathrm{F}}\left(\tau_{2}\right)$ for (c) $n_{\mathrm{iL}}$ and (d) $n_{\mathrm{iH}}$ are shown.}
	\end{figure}

	Figure\;5(a) presents the temporal evolutions of $T_{\mathrm{e}}$ and $T_{\eta}$ of the heavily doped graphene by its photoexcitation with $F_{\text {ab }}=0.04$ and $0.13\, \upmu \mathrm{Jcm}^{-2}$. 
	The maximum values of $T_{\mathrm{e}}$ and $T_{\eta}$ for $F_{\text {ab }}=0.04$ and $0.13\;\upmu \mathrm{Jcm}^{-2}$ are approximately 700 and $1,200\,\mathrm{K}$, which are smaller than those of the undoped graphene because of the larger specific heat capacity, $C$, and total balance of the energy exchange rate, $R_{\eta}^{\mathrm{Net}} \hbar \omega_{\eta}$, of the heavily doped graphene (see Appendix C). 
	The relaxation curves of $T_{\mathrm{e}}$ exhibit double-exponential decays with $\tau_{\mathrm{T} 1}=0.28$ and $0.27\,\mathrm{ps}$ and $\tau_{\mathrm{T} 2}=2.7$ and $2.6\,\mathrm{ps}$ for $F_{\mathrm{ab}}=0.04$ and $0.13\,\upmu \mathrm{J} \mathrm{cm}^{-2}$, respectively. 

	Figures\:6(a) and (b) show $\sigma\left(\omega, \tau_{1}\right)$ of the heavily doped graphene with $n_{\mathrm{iL}}$ and $n_{\mathrm{iH}}$ calculated using the THz-probe pulse with $2 \tau_{\mathrm{p}}=300\,\mathrm{fs}$, respectively. $\sigma\left(\omega, \tau_{1}\right)$ at equilibrium show the simple Drude-like frequency dependence and are larger than those of undoped graphene, owing to large carrier concentrations. 
	By the photoexcitation, both $\sigma\left(\omega, \tau_{1}\right)$ slightly decrease, indicating a negative photoconductivity as reported in past studies\cite{Jnawali2013b, Frenzel2014d, Shi2014a, Kar2014a, Jensen2014b}.
	As seen in Figs\;6(c)--(f), the frequency dependencies of $D\left(\omega, \tau_{1}\right)$ and $\Gamma\left(\omega, \tau_{1}\right)$ at the hot-carrier state show the similar behaviors for $n_{\mathrm{iL}}$ and $n_{\mathrm{iH}}$, but they are different from those of undoped graphene. Because the strong carrier-screening effect in highly doped graphene weakens the charged impurity scattering considerably, this indicates that the optical phonon scattering is dominant at the hot carrier state in heavily doped graphene. In fact, the $\Gamma\left(\omega, \tau_{1}\right)$ value at the hot-carrier state slightly increases. In contrast with undoped graphene, $D\left(\omega, \tau_{1}\right)$ decreases with increasing $T_{\mathrm{e}}$ after photoexcitation and takes the minimum around 2,000 K, which can be attributed to the unique behaviors of $D\left(T_{\mathrm{e}}\right)$ of 2D-MDF in heavily doped graphene  (see Fig. A.1(a)). 
	Therefore, both $D\left(\omega, \tau_{1}\right)$ and $\Gamma\left(\omega, \tau_{1}\right)$ contribute to the negative photoconductivity of the hot-carrier state in heavily doped graphene. 
	
	Figure 7 shows the temporal waveforms of THz-field-induced current densities, $J\left(\tau_{2}\right)$ and $J_{\mathrm{F}}\left(\tau_{2}\right)$, in heavily doped graphene. The $\Delta J_{\mathrm{F}}$ for $n_{\mathrm{iL}}$ and $n_{\mathrm{iH}}$ are similar and sufficiently small. As a result,  $\sigma_{\mathrm{F1}}\left(\omega, \tau_{1}\right)$ well-agree with the $\sigma_{1}\left(\omega, \tau_{1}\right)$ of hot carriers. However, the waveforms of $\Delta J_{\mathrm{F}}$ of heavily doped graphene are inverted from those of undoped ones, making $\sigma_{\mathrm{2}}\left(\omega, \tau_{1}\right)$ smaller than $\sigma_{\mathrm{F2}}\left(\omega, \tau_{1}\right)$ in the low-frequency region where $D\left(\omega, \tau_{1}\right)$ and $\Gamma\left(\omega, \tau_{1}\right)$ decrease with increasing frequency.  
	The results of RTA calculations well-reproduce the magnitude and frequency dependence of $\sigma\left(\omega, \tau_{1}\right)$ at equilibrium. However, they fail for the hot-carrier state. The frequency dependence of the $D\left(\omega, \tau_{1}\right)$ and $\Gamma\left(\omega, \tau_{1}\right)$ significantly differ from those of the iterative method with both varied and fixed temperatures, which can be attributed to the relaxation rate, $\tau_{\mathrm{op}}$, by the optical phonon, which uses the Maxwell--Boltzmann distribution instead of the 
	FD distribution in RTA calculation  (see Appendix B). Note that, at the equilibrium state ($T_{\mathrm{e}}=295\,\mathrm{K}$), optical phonon scattering is sufficiently weak, such that the $D\left(\omega, \tau_{1}\right)$ and $\Gamma\left(\omega, \tau_{1}\right)$ show almost constant values. 
			
	The corresponding temporal evolutions of $-\Delta E_{t}\left(\tau_{1}\right) / E_{0}$ of heavily doped graphene calculated using the THz-probe pulse with $2\tau_{\mathrm{p}}=300\,\mathrm{fs}$ are shown in Fig.\;4(b). They exhibit double exponential relaxation curves with $\tau_{\mathrm{THz} 1}=0.32$ and $0.28\,\mathrm{ps},$ and $\tau_{\mathrm{TH} 2}=2.0$ and $1.9\,\mathrm{ps}$ for $n_{\mathrm{iL}}$ and $\tau_{\mathrm{THz} 1}=0.32$ and $0.28\,\mathrm{ps},$ and $\tau_{\mathrm{THz} 2}=2.0$ and $1.9\,\mathrm{ps}$ for $n_{\mathrm{iH}}$, respectively. The decay times, $\tau_{\mathrm{THz} 1}$ and $\tau_{\mathrm{THz} 2}$, of $-\Delta E_{t}\left(\tau_{1}\right) / E_{0}$ roughly reflect the relaxation times, $\tau_{\mathrm{T} 1}$ and $\tau_{\mathrm{T} 2}$, of the hot carriers in heavily doped graphene.

\section{Discussion and Conclusion}
	In Section III, we presented the numerical results of the intraband optical conductivity of hot carriers in undoped and heavily doped graphene after photoexcitation, considering the intrinsic and extrinsic carrier scattering mechanisms that exhibits positive and negative photoconductivity, depending on the Fermi energy. In undoped graphene, the large positive photoconductivity arises from the positive change of the hot-carrier population surpassing the negative contribution by the enhanced carrier scattering by the optical phonon. 
	The charged impurity scattering decreases because of the enhanced carrier-screening effect, resulting in the larger positive photoconductivity of hot carriers in undoped graphene.
	In heavily doped graphene, the Drude weight decreases with $T_{\mathrm{e}}$, owing to the unique temperature dependence of 2D-MDF. Because charged impurity scattering is strongly suppressed by the carrier screening effect, both the reduction of Drude weight and the increased optical phonon scattering contribute to the negative photoconductivity in the hot-carrier state. 
	
	Most experimental studies, in fact, have been conducted using graphene at intermediate doping levels. Here, we present the theoretical results on lightly $p$-doped graphene and discuss the origin of their unique behaviors related to $\sigma\left(\omega, \tau_{1}\right)$ and $-\Delta E_{t}\left(\tau_{1}\right) / E_{0}$.
	\begin{figure}[b]
		\centering
		\includegraphics[width=5.5cm, bb=0 0 283 571]{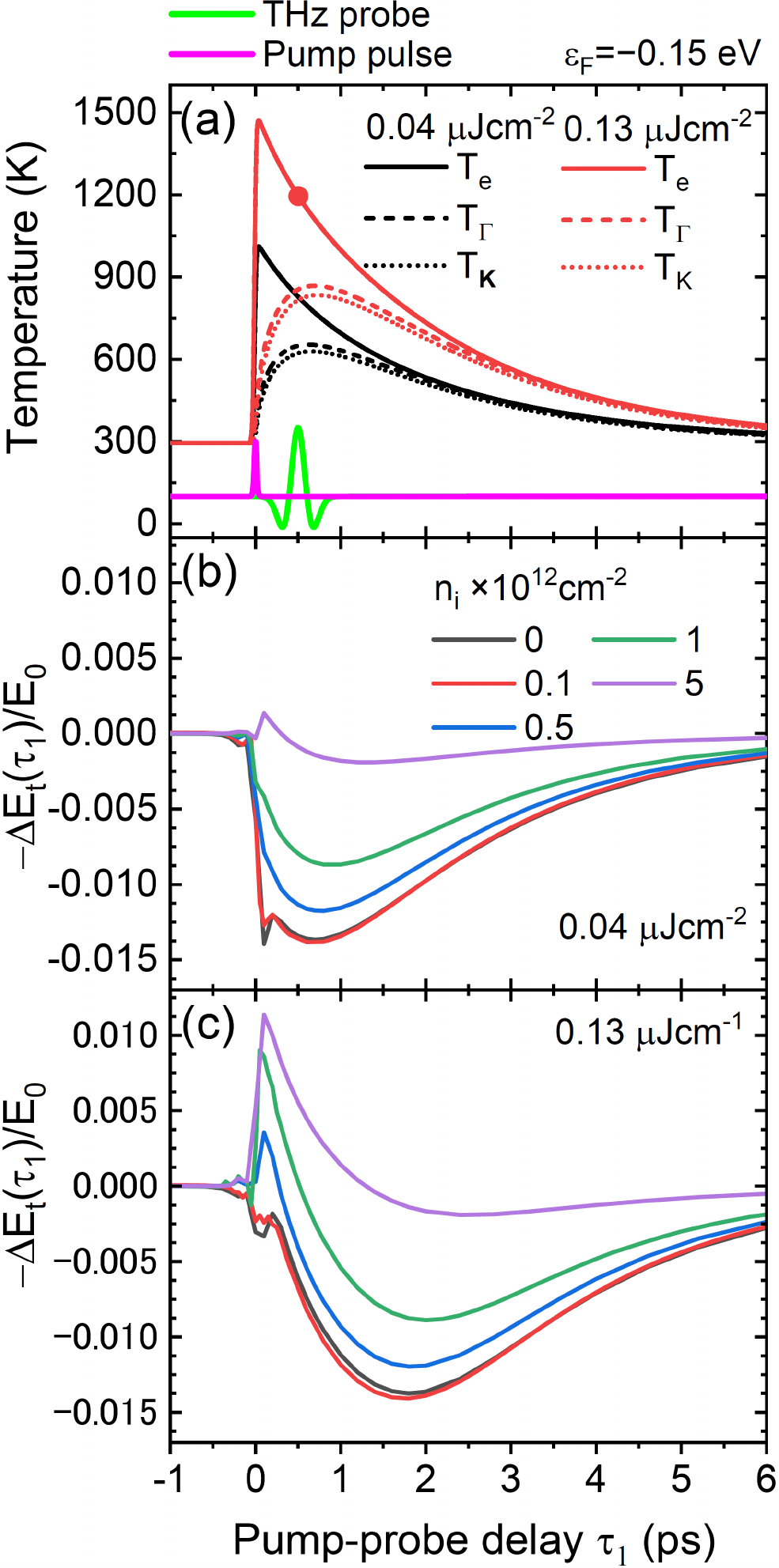}
		\caption{\label{fig9} (a) Temporal evolutions of $T_{\mathrm{e}}$ and $T_{\eta}$ of lightly doped graphene for $F_{\mathrm{ab}}=0.04$ and $0.13\,\mathrm{\upmu Jcm}^{-2}$. Temporal evolutions of $-\Delta E_{\mathrm{t}}\left(\tau_{1}\right) / E_{0}$  with various charged impurity concentration, 
			$n_{\mathrm{i}}=0,\;0.1,\;0.5,\;1.0,\;5.0 \times10^{12}\,\mathrm{cm}^{-2}$, for $F_{\mathrm{ab}}=$ (b) 0.04 and (c) $0.13\,\mathrm{\upmu Jcm}^{-2}$, calculated using the THz-probe pulse with $2 \tau_{\mathrm{p}}=300\,\mathrm{fs}$.}
	\end{figure}
	Figure\;8(a) presents the temporal evolutions of $T_{\mathrm{e}}$ and $T_{\eta}$ in lightly doped graphene with $\varepsilon_{\mathrm{F}}=-0.15\,\mathrm{eV}$ . The corresponding hole concentration at $T=0\,\mathrm{K}$ is $n_{\mathrm{c}}=1.6 \times 10^{12}\,\mathrm{cm}^{-2}$. 
	The $T_{\mathrm{e}}$ curves with $F_{\mathrm{ab}}=0.04$ and $0.13\,\upmu \mathrm{J} / \mathrm{cm}^{-2}$ exhibit double exponential decays with $\tau_{\mathrm{T} 1}=0.55$ and $0.21\,\mathrm{ps}$ and $\tau_{\mathrm{T} 2}=2.1$ and $2.1\,\mathrm{ps}$ for $F_{\mathrm{ab}}=0.04$ and $0.13\,\upmu \mathrm{J} \mathrm{cm}^{-2}$, respectively.
	$T_{\mathrm{e}}$ increases to 1,000 and $1,500\,\mathrm{K}$, respectively, which is close to the values of the undoped graphene owing to the similar specific heat capacity, $C$, and total balance of the energy exchange rate, $R_{\eta}^{\mathrm{Net}} \hbar \omega_{\eta}$ (see Appendix C). 
	Figure 9(a) and (b) show the  $\sigma\left(\omega, \tau_{1}\right)$ of the lightly doped graphene with $n_{\mathrm{iL}}=0.1 \times 10^{12}\,\mathrm{cm}^{-2}$ and $n_{\mathrm{iH}}=1.0 \times 10^{12}\,\mathrm{cm}^{-2}$ calculated using the THz-probe pulse with $2 \tau_{\mathrm{p}}=300\,\mathrm{fs}$, respectively, for $F_{\mathrm{ab}}=0.13\,\upmu \mathrm{J} / \mathrm{cm}^{-2}$. The corresponding DC conductivities at $T_0=295\,\mathrm{K}$ are $\sigma_{\mathrm{DC}}=21.5 G_{0}$ and $7.8 G_{0}$. The $\sigma\left(\omega, \tau_{1}\right)$ with $n_{\mathrm{iH}}$ at equilibrium is substantially suppressed, compared to that of $n_{\mathrm{iL}}$ by the larger charged impurity scattering caused by weaker carrier screening versus the heavily doped graphene.
	\begin{figure}[b]
		\centering		\includegraphics[width=8.6cm, bb=0 0 283 335]{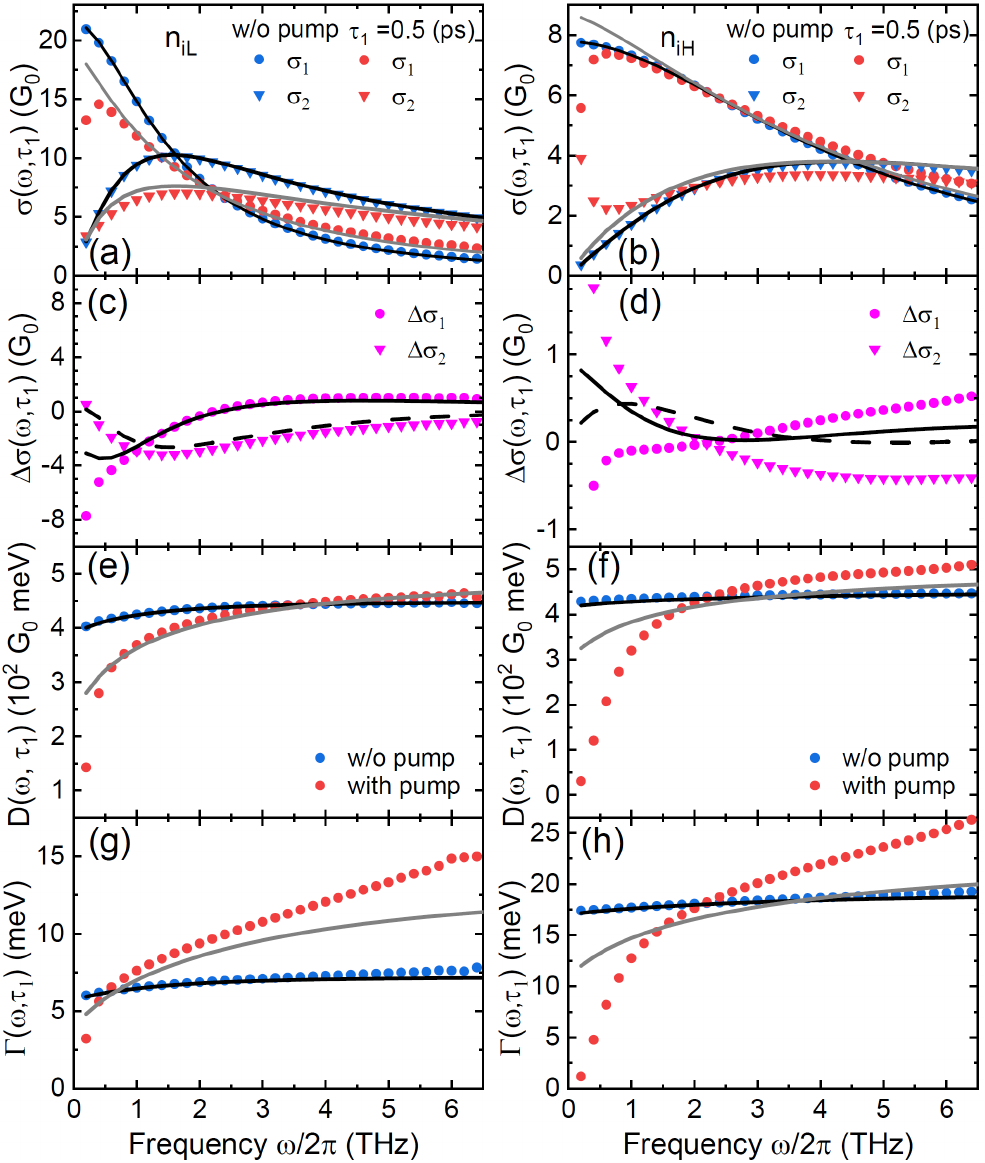}
		\caption{\label{fig9} Simulation results of lightly doped graphene calculated using the THz-probe pulse with $2 \tau_{\mathrm{p}}=300\,\mathrm{fs}$. The $\sigma\left(\omega, \tau_{1}\right)$ for (a) $n_{\mathrm{iL}}$ and (b) $n_{\mathrm{iH}}$ with $F_{\mathrm{ab}}=0.13\,\upmu \mathrm{J} \mathrm{cm}^{-2}$ at $\tau_{1}=0.5\,\mathrm{ps}$, indicated by the filled solid circle in Fig. 8(a). The $\Delta \sigma\left(\omega, \tau_{1}\right)$ for (c) $n_{\mathrm{iL}}$ and (d) $n_{\mathrm{iH}}$. Frequency dependence of $D\left(\omega, \tau_{1}\right)$ for (e) $n_{\mathrm{iL}}$ and (f) $n_{\mathrm{iH}}$. Frequency dependence of $\Gamma\left(\omega, \tau_{1}\right)$ for (g)) $n_{\mathrm{iL}}$ and (h)  $n_{\mathrm{iH}}$. The blue and red symbols are calculated by the iterative method with varied temperatures. 
		The black and gray lines reflect the RTA formula of Eq. (B10). The RTA calculations well-agree with the iterative method at equilibrium state, but not at hot-carrier state.}
	\end{figure}
	The effect of photoexcitation on $\sigma\left(\omega, \tau_{1}\right)$ of lightly doped graphene are different from both undoped and heavily doped graphenes.
	The $\sigma\left(\omega, \tau_{1}\right)$ of the hot carriers for both $n_{\mathrm{iL}}$ and $n_{\mathrm{iH}}$ show negative and positive photoconductivity, depending on the frequency. As shown in Figs. 9(e)--(h), the frequency dependencies of $D\left(\omega, \tau_{1}\right)$ and $\Gamma\left(\omega, \tau_{1}\right)$ of hot carriers are similar to those of undoped graphene. Their positive slopes become stronger on the low-frequency side, resulting in the non-Drude behaviors of $\sigma\left(\omega, \tau_{1}\right)$ in the low frequency region, reflecting the energy dependence of the scattering rate. Because the changes of $D\left(\omega, \tau_{1}\right)$ in lightly doped graphenes caused by the photoexcitation are considerably smaller than undoped graphene, the $\Gamma\left(\omega, \tau_{1}\right)$  differentiates the behavior of $\sigma\left(\omega, \tau_{1}\right)$ between $n_{\mathrm{iL}}$ and $n_{\mathrm{iH}}$. 
	The increased optical phonon and decreased charged impurity scattering for $n_{\mathrm{iH}}$ cancel each other out, resulting in a smaller positive change of the $\Gamma\left(\omega, \tau_{1}\right)$ and $\Delta \sigma\left(\omega, \tau_{1}\right)$ than for $n_{\mathrm{iL}}$.
	\begin{figure}
		\centering
		\includegraphics[width=8.5cm, bb=0 0 283 195]{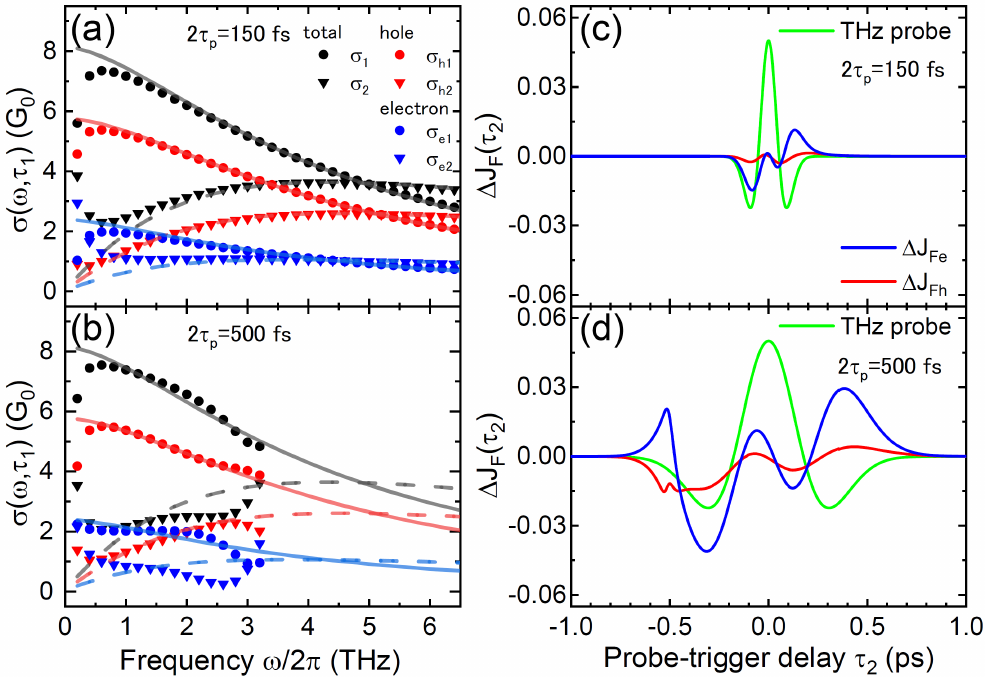}
		\caption{\label{figD1} The $\sigma\left(\omega, \tau_{1}\right)$, $\sigma_{\mathrm{e}}\left(\omega, \tau_{1}\right)$ for electrons and $\sigma_{\mathrm{h}}\left(\omega, \tau_{1}\right)$ for holes of lightly doped graphene with $n_{\mathrm{iH}}$ for $F_{\mathrm{ab}}=0.13\,\upmu \mathrm{J} \mathrm{cm}^{-2}$ at $\tau_{1}=0.5\,\mathrm{ps}$, calculated using THz probe with (a) $2 \tau_{\mathrm{p}}=150$ and (b) $500\,\mathrm{fs}$, respectively, where $\sigma_{\mathrm{e}}\left(\omega, \tau_{1}\right)=\sigma_{\mathrm{e}1}\left(\omega, \tau_{1}\right)+ i \sigma_{\mathrm{e}2}\left(\omega, \tau_{1}\right)$ and $\sigma_{\mathrm{h}}\left(\omega, \tau_{1}\right)=\sigma_{\mathrm{h}1}\left(\omega, \tau_{1}\right)+ i \sigma_{\mathrm{h}2}\left(\omega, \tau_{1}\right)$.
		The gray, red and blue lines are calculated by the iterative method with fixed temperatures. For $2 \tau_{\mathrm{p}}=500\,\mathrm{fs}$, $\sigma\left(\omega, \tau_{1}\right)$ with $\omega /2 \pi > 3.2\,\mathrm{THz}$ is not shown because the limited bandwidth of the THz probe causes the numerical errors. Temporal waveforms of the THz-probe pulse and the normalized difference of electron and hole current densities, $\Delta J_{\mathrm{Fe}}$ and $\Delta J_{\mathrm{Fh}}$, for (c) $2 \tau_{\mathrm{p}}=150$ and (d) $500\,\mathrm{fs}$.}
	\end{figure}

	From the results of undoped and heavily doped graphene, the origin of the non-Drude behavior can be attributed to the temporal variation of the hot-carrier distribution and carrier scattering in addition to the energy dependence of the momentum relaxation rate. Furthermore, the non-Drude behaviors were remarkable by the presence of charged impurities. The frequency dependence of the non-Drude type photoconductivity of lightly doped graphene is similar to that of undoped graphene, but it is different from that of heavily doped graphene. 
	To gain a deeper insight, we present in Fig. 10 a comparison of $\sigma\left(\omega, \tau_{1}\right)$ of lightly doped graphene with $n_{\mathrm{iH}}$, calculated using the THz probe with $2\tau_{\mathrm{p}}=150$ and $500\,\mathrm{fs}$. It is clearly seen that the $\sigma\left(\omega, \tau_{1}\right)$ at the hot-carrier state strongly depends on the waveform of the THz probe.  The difference between $\sigma\left(\omega, \tau_{1}\right)$ and $\sigma_{\mathrm{F}}\left(\omega, \tau_{1}\right)$ for $2\tau_{\mathrm{p}}=150\,\mathrm{fs}$ is substantially smaller than that for $2\tau_{\mathrm{p}}=500\,\mathrm{fs}$. This supports the notion that the larger temporal variations of carrier distribution and scattering rate during the THz probing cause more prominent non-Drude behaviors. In addition, it is seen that the deviation of $\sigma_{\mathrm{2}}\left(\omega, \tau_{1}\right)$ from $\sigma_{\mathrm{F2}}\left(\omega, \tau_{1}\right)$ for the electron current is significantly larger than that for the hole current, whereas the $\sigma_{\mathrm{1}}\left(\omega, \tau_{1}\right)$ of the electron and hole current shows 
	similar frequency dependencies. The THz-probe dependence of the normalized difference of current density shown in Figs. 10 (c) and (d) indicates that the $\Delta J_{\mathrm{Fe}}$ of the electron current is larger than $\Delta J_{\mathrm{Fh}}$ of the hole current, although electrons are minority charge carriers in the p-type doped graphene. This is because the population of the minority charge carriers strongly depends on $T_{\mathrm{e}}$ and the minor carrier distribution change greatly after photoexcitation. Furthermore, a comparison of the energy dependence of the relaxation rate is shown in Fig. B.1 in Appendix B revealing that the charged impurity scattering is most sensitive to the change of carrier distribution for $|\varepsilon_{\lambda \textbf{\textit{k}}}|<0.2\,\mathrm{eV}$ where optical phonon emission is suppressed  and that it highly contributes to the emergence of the non-Drude behaviors of $\sigma_{\mathrm{e2}}\left(\omega, \tau_{1}\right)$ for the electron current in undoped and lightly doped graphene. This indicates that, with an increasing carrier-doping level, the minority carrier distribution and their scattering rates by charged impurity at the hot carrier state become negligible resulting in the disappearance of non-Drude behaviors in $\sigma_{\mathrm{2}}\left(\omega, \tau_{1}\right)$, as seen in heavily doped graphene.

	The $-\Delta E_{\mathrm{t}}\left(\tau_{1}\right) / E_{0}$ of the lightly doped graphene presented in Fig. 8(b) exhibits unique behaviors depending on $F_{\mathrm{ab}}$ and $n_{\mathrm{i}}$. For a weak-pump condition, $F_{\mathrm{ab}}=0.04\,\mathrm{\upmu Jcm}^{-2}$, $T_{\mathrm{e}}$ varies between $295\,\mathrm{K}$ and $1050\,\mathrm{K}$  and the $-\Delta E_{\mathrm{t}}\left(\tau_{1}\right) / E_{0}$ without charged impurities shows negative photoconductivity, which first exhibits a sharp peak followed by a second negative broad peak. The first peak corresponds to the reduction of the Drude weight, which decreases slightly with an increasing $T_{\mathrm{e}}$ below $1,000\,\mathrm{K}$. The second peak corresponds to an increased scattering rate caused by the hot optical phonon. The heights of the first and second peaks are comparable, indicating similar contributions of Drude weight and optical phonon scattering to the negative photoconductivity. The negative photoconductivity decreases clearly with increasing charged impurities, indicating that the dominant scattering mechanism changed from the optical phonon to charged impurities, because the momentum relaxation rate of hot carriers by the charged impurity scattering changes little in the temperature range between $295\,\mathrm{K}$ and $1,050\,\mathrm{K}$.
	For strong-pump condition $F_{\mathrm{ab}}=0.13\,\mathrm{\upmu Jcm}^{-2}$, $T_{\mathrm{e}}$ varies between $295\,\mathrm{K}$ and $1,500\,\mathrm{K}$. In this case, the $-\Delta E_{\mathrm{t}}\left(\tau_{1}\right) / E_{0}$ lacking charged impurities shows a large negative broad peak caused by the increased optical phonon scattering. The small dip at $\tau_{1}=0.2\,\mathrm{ps}$ corresponds to the positive change of the Drude weight above $T_{\mathrm{e}}=1,000\,\mathrm{K}$ and changes to the positive peak with increasing the charged impurity, because the increase of the carrier screening effect at $T_{\mathrm{e}} \geq1,000\,\mathrm{K}$ decreases the charged impurity scattering. Therefore, the crossover from negative photoconductivity to a positive one without changing the carrier concentration also indicates a change to the dominant scattering mechanism from optical phonons to charged impurities.
	
	The crossover from negative to positive photoconductivity was experimentally observed by Docherty $et\;al.$ \cite{Docherty2012e}. The $-\Delta E_{\mathrm{t}}\left(\tau_{1}\right) / E_{0}$ of doped graphene having $\varepsilon_{\mathrm{F}}=-0.3\,\mathrm{eV}$ on a quartz substrate exhibited a strong dependence on different atmospheric environments.  Although the negative photoconductivity was observed with the presence of $\mathrm{O}_{2}$, air, or $\mathrm{N}_{2}$, a positive photoconductivity appeared with the vacuum environment. Additionally, the photoconductivity spectra under the gas atmosphere showed a Lorentzian form having a negative amplitude. The $\Delta \sigma_{1}$ exhibited a negative minimum value around $\omega_{0} / 2 \pi=1.8\,\mathrm{THz}$, and the $\Delta \sigma_{2}$ 
	decreased with the frequency, thereby exhibiting a zero crossing. The origin of negative amplitude in the Lorentzian model was suggested to be caused by the stimulated emission due to the population inversion in the photoexcited graphene with a small band gap opened by 
	gas-molecule adsorption. However, this suggestion has been debated, because the time- and angle-resolved photoemission experiment by Gierz $et\;al.$\cite{Gierz2013c} demonstrated a time duration of population inversion within $130\,\mathrm{fs}$ after photoexcitation, owing to a strong Auger recombination in the doped graphene.
	Based on our results, the origin of the reported crossover of photoconductivity is suggested to have been caused by the increase of the charged impurity scattering, owing to the changes from the gas atmosphere to a vacuum. Moreover, the $\Delta \sigma\left(\omega, \tau_{1}\right)$ with $n_{\mathrm{iH}}$ by the iterative method shown in Fig. 9 (d) has a distribution close to Lorentzian model with a negative amplitude in the low frequency region, $0.4<\omega /2 \pi <2.5\,\mathrm{THz}$, which is almost equal to the observed frequency region. Because the Fermi energy changed only 3 $\%$ by the atmosphere \cite{Docherty2012e}, it rules out the crossover mechanism from the doped graphene to the undoped one \cite{Frenzel2014d, Shi2014a} or by the change of carrier-screening effect. Thus, molecular gas adsorption may have played a role of the reduction of charged impurity scattering by increasing the background dielectric constant and/or reducing the effective charged impurity concentration.

	In summary, we studied the intraband optical conductivity of hot carriers at quasi-equilibrium in photoexcited graphene based on a combination of the iterative solutions for the BTE and comprehensive temperature model. 
	The proposed method enables us to consider the extrinsic effects such as charged impurity scattering and the intrinsic effect of optical phonon scattering on the hot-carrier THz conductivity, having a reduced computational cost compared with the full BTE solution. In the examples of photoexcited graphene having different Fermi energies, we demonstrated the temporal evolution and frequency dependence of the photoconductivity of hot carriers 
	exhibiting a strong dependence on the temporal variation of carrier distributions and scattering rates. 
	Our method provides a quantitative analysis of hot-carrier THz conductivity with intrinsic and extrinsic microscopic parameters, such as electron--phonon coupling and charged impurity concentration, which are important to the development of future graphene optoelectronic devices. 
	
	\begin{acknowledgments}
		This work was supported by the JSPS KAKENHI (19H01905) and Research Foundation for Opto-Science and Technology.
		
	\end{acknowledgments}

	\appendix
	\counterwithin{figure}{section}
	
	\section{Temperature dependence of chemical potential and Drude weight} 
	In $n$-type doped graphene, the carrier concentration $n_{0}$ at $T_{\mathrm{e}}=0\,K$ in the conduction band is expressed as a function of the Fermi energy, $\varepsilon_{\mathrm{F}}$:
		\begin{equation}
	n_{0}=\int_{0}^{\varepsilon_{\mathrm{F}}} N\left(\varepsilon_{1 \textbf{\textit{k}}}\right) d \varepsilon_{1\textbf{\textit{k}}}=\int_{0}^{\varepsilon_{\mathrm{F}}} N\left(\varepsilon_{1 \textbf{\textit{k}}}\right) d \varepsilon_{1 \textbf{\textit{k}}}.
	\end{equation}
	At a finite $T_{\mathrm{e}}$, the total carrier concentration $n\left(T_{\mathrm{e}}\right)$ in the conduction band is given by
	\begin{equation}
	n\left(T_{\mathrm{e}}\right)=\int_{0}^{\infty} N\left(\varepsilon_{1 \textbf{\textit{k}}}\right) \frac{1}{\mathrm{e}^{\left[\left(\varepsilon_{1 \textbf{\textit{k}}}-\mu\left(T_{\mathrm{e}}\right)\right) / k_{\mathrm{B}} T_{\mathrm{e}}\right]}+1} d \varepsilon_{1 \textbf{\textit{k}}}.
	\end{equation}
	Here, $\mu\left(T_{\mathrm{e}}\right)$ is the finite temperature chemical potential, and $n\left(T_{\mathrm{e}}\right)$ is expressed as the sum of $n_{0}$ and the thermally activated carrier concentration $n_{\mathrm{T}}$ from the valence band.
	\begin{equation}\begin{aligned}
	n\left(T_{\mathrm{e}}\right)&=n_{0}+n_{\mathrm{T}}\\
	&=n_{0}+\int_{0}^{\infty} N\left(\varepsilon_{1\textbf{\textit{k}}}\right) \frac{1}{\mathrm{e}^{\left[\left(\varepsilon_{1 \textbf{\textit{k}}}+\mu\left(T_{\mathrm{e}}\right)\right) / k_{\mathrm{B}} T_{e}\right]}+1} d \varepsilon_{1 \textbf{\textit{k}}}.
	\end{aligned}\end{equation}
	As a result, $n_{0}$ is expressed as a function of the temperature-dependent chemical potential, $\mu\left(T_{\mathrm{e}}\right)$, as
	\begin{equation}\begin{aligned}
	n_{0} =\int_{0}^{\infty} N\left(\varepsilon_{1 \textbf{\textit{k}}}\right)&\left\{\frac{1}{\mathrm{e}^{\left[\left(\varepsilon_{1 \textbf{\textit{k}}}-\mu\left(T_{\mathrm{e}}\right)\right) / k_{\mathrm{B}} T_{\mathrm{e}}\right]}+1}\right.\\
	&\left.-\frac{1}{\mathrm{e}^{\left[\left(\varepsilon_{1 \textbf{\textit{k}}}+\mu\left(T_{\mathrm{e}}\right)\right) / k_{\mathrm{B}} T_{\mathrm{e}}\right]}+1}\right\} d \varepsilon_{1\textbf{\textit{k}}}.
	\end{aligned}\end{equation}
	Because $n_{0}$ is constant and given by Eq.\;(A1), the temperature dependence of the chemical potential can be obtained by numerically inverting Eq.\;(A4) \cite{Frenzel2014d}. 
	Figure\;A.1(a) presents the chemical potential of the charge carriers in graphene as a function of temperature. 
	In $p$-type doped graphene, the temperature dependence of the chemical potential can be obtained similarly. 
	In the Boltzmann theory, assuming a constant carrier relaxation rate, the Drude weight of the 2D-MDF exhibits a unique temperature dependence as given by \cite{Muller2009, Gusynin2009, Wagner2014, Frenzel2014d}.
	\begin{equation}
	D\left(T_{\mathrm{e}}\right)=\frac{2 e^{2}}{\hbar^{2}} k_{\mathrm{B}} T_{\mathrm{e}} \ln \left[2 \cosh \left(\frac{\mu\left(T_{\mathrm{e}}\right)}{2 k_{\mathrm{B}} T_{\mathrm{e}}}\right)\right].
	\end{equation}
	The temperature dependencies of $D\left(T_{\mathrm{e}}\right)$ for different Fermi energies are shown in Fig.\;A.1(b).
	
	\begin{figure}
		\centering
		\includegraphics[width=8.6cm, bb=0 0 241 104]{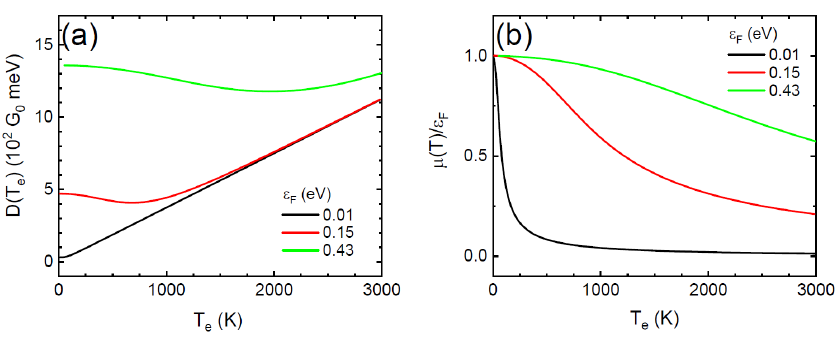}
		\caption{\label{figA1} Temperature dependence of the chemical potential, $\mu\left(T_{\mathrm{e}}\right)$, and Drude weight, $D\left(T_{\mathrm{e}}\right)$, of graphene with $\left|\varepsilon_{\mathrm{F}}\right|=0.01, 0.15$ and $0.43\,\mathrm{eV}$.}
	\end{figure}

	\section{Calculation of relaxation rate using RTA}
	We demonstrate the numerical results of the calculation of the inelastic scattering by an intrinsic non-polar optical phonon in monolayer graphene. 
	Figures\;B.1(a)--(c) present $S_{\lambda}^{\text{out}}$ as a function of the carrier energy for graphene having $|\varepsilon_{\mathrm{F}}|=0.01, 0.15, 0.43\,\mathrm{eV}$, respectively, calculated using Eqs.\;(7), (8), and (12). 
	For comparison, we show the momentum relaxation rate calculated using RTA \cite{Lundstrom2009a, Hamaguchi2010a}. 
	In the case of nondegenerate semiconductors, the Maxwell--Boltzmann distribution $n_{\mathrm{MB}}=1/\mathrm{Exp}[(\varepsilon-\mu)/k_{\mathrm{B}}T]$ with $\mu=0\,\mathrm{eV}$ is used instead of the FD distribution for the calculation of the momentum relaxation rate by non-polar optical phonons with mode $\eta$, which is given by
	\begin{equation}
	\tau_{\eta \pm}^{-1}\left(\varepsilon_{\lambda \textbf{\textit{k}}}\right)=\frac{D_{\mathrm{\eta}}^{2}}{\rho \omega_{\mathrm{\eta}} \hbar^{2} v_{\mathrm{F}}^{2}}\left(\varepsilon_{\lambda \textbf{\textit{k}}} \mp \hbar \omega_{\mathrm{\eta}}\right)\left(n_{\mathrm{\eta}}+\frac{1}{2} \pm \frac{1}{2}\right),
	\end{equation}
	where the top sign applies to phonon emission, and the bottom sign applies to phonon absorption. 
	Furthermore, $D_{\eta}$ is the deformational potential of the optical phonon mode, $\eta$. We use $D_{\eta}^{2}=\left\langle D_{\eta}^{2}\right\rangle_{F}$ for the RTA calculation. 
	The total optical phonon scattering by RTA is then expressed by
	\begin{equation}
	\tau_{\mathrm{op}}^{-1}\left(\varepsilon_{\lambda \textbf{\textit{k}}}\right)=\sum_{\eta \pm} \tau_{\eta \pm}^{-1}\left(\varepsilon_{\lambda \textbf{\textit{k}}}\right).
	\end{equation}
	Figure\;B.1(d) presents the energy dependence of $\tau_{\mathrm{op}}^{-1}$, which is larger than $S_{\lambda}^{\text{out}}$. This is because the scattering-angle dependence is not included for the calculation of $\tau_{\mathrm{op}}^{-1}$.
	
	Moreover, we demonstrate elastic or quasielastic carrier scattering by charged impurities \cite{Adam2007b, Tan2007b, Ando2006a, Chen2008h, Chen2008f, Hwang2007k}, weak scatterers \cite{Lin2014d, Stauber2007b, Adam2008b, Katsnelson2008a, Morozov2008b, Dean2010b, Pachoud2010b, Yan2011b} and acoustic phonons \cite{Tan2007b, Chen2008h, Chen2008f, Hwang2007k, Morozov2008b, Dean2010b, Perebeinos2010b, Tanabe2011a, Zou2010, Hwang2008g, Castro2010a} considered in the iterative calculation of the THz conductivity. 
	For elastic scattering, the RTA under the low-field condition is valid \cite{Lundstrom2009a, Hamaguchi2010a}, and the momentum relaxation rate in graphene is equal to the relaxation rate of the distribution function. 
		
	Charged impurity scattering plays an important role in the carrier transport and intraband conductivity of undoped or lightly doped graphene on a substrate. The carrier scattering by charged impurities at the graphene-substrate interface limits the carrier mobility significantly. The relaxation rate, $\tau_{\mathrm{i}}^{-1}\left(\varepsilon_{\lambda \mathrm{\textbf{\textit{k}}}}\right)$, for the charged impurities under RTA is expressed by
	\begin{equation}\begin{aligned}
	\tau_{\mathrm{i}}^{-1}\left(\varepsilon_{\lambda \textbf{\textit{k}}}\right)=& \frac{\pi n_{\mathrm{i}}}{\hbar} \int \frac{d^{2} \boldsymbol{k}^{\prime}}{(2 \pi)^{2}}\left|\frac{v_{\mathrm{i}}(p)}{\epsilon\left(p, T_{\mathrm{e}}\right)}\right|^{2} \\
	& \times\left(1-\cos ^{2} \theta_{\textbf{\textit{k}}\textbf{\textit{k}}^{\prime}}\right) \delta\left(\left|\varepsilon_{\lambda \textbf{\textit{k}}}\right|-\left|\varepsilon_{\lambda \textbf{\textit{k}}^{\prime}}\right|\right).
	\end{aligned}\end{equation}
	Here, $n_{\mathrm{i}}$ is the charged impurity concentration, $p=\left|\mathbf{k}-\mathbf{k}^{\prime}\right|$ is the scattering wave vector, and $v_{\mathrm{i}}(p)=e^{2} / 2 \epsilon_{\mathrm{ave}} p$ is the Fourier transform of the 2D Coulomb potential in an effective background lattice permittivity, $\epsilon_{\text {ave }}=\left(1+\epsilon_{\mathrm{s}}\right)\epsilon_{0} / 2$, given by the average static dielectric constant of the vacuum and substrate. Here, $\epsilon_{0}$ is the vacuum permittivity. Furthermore, $\epsilon\left(p, T_{\mathrm{e}}\right)$ is the static electronic dielectric function of graphene calculated by the random-phase approximation, and it is responsible for the screening effect.
	\begin{equation}
	\epsilon\left(p, T_{\mathrm{e}}\right)=1+\frac{e^{2}}{2 \epsilon_{\mathrm{ave}} p} \Pi\left(p, T_{\mathrm{e}}\right),
	\end{equation}
	where $\Pi\left(p, T_{\mathrm{e}}\right)$ is the graphene irreducible finite-temperature polarizability function expressed by
	\begin{equation}
	\Pi\left(p, T_{\mathrm{e}}\right)=-\frac{\mathrm{g}}{A} \sum_{\textbf{\textit{k}} \lambda \lambda^{\prime}} \frac{f_{\lambda}(\boldsymbol{k})-f_{\lambda^{\prime}}\left(\boldsymbol{k}^{\prime}\right)}{\varepsilon_{\lambda \textbf{\textit{k}}}-\varepsilon_{\lambda^{\prime} \textbf{\textit{k}}^{\prime}}}\left(1+\lambda \lambda^{\prime} \cos \theta_{\boldsymbol{k} \boldsymbol{k}^{\prime}}\right),
	\end{equation}
	in which $\mathrm{g}=4$ is the spin and valley degeneracy factor, $A$ is the area of the system, and $f_{\lambda}(\textbf{\textit{\textit{k}}})$ is the carrier distribution function. Moreover, $f_{\lambda}(\boldsymbol{k}) \simeq f_{0}\left(\varepsilon_{\lambda \textbf{\textit{k}}}\right)$ is used in the calculation of $\Pi\left(p, T_{\mathrm{e}}\right)$, because we consider the small perturbed distribution, $g\left(\varepsilon_{\lambda \textbf{\textit{k}}}\right)$. The temperature dependence of the charged impurity scattering arises from the temperature-dependent carrier screening of the Coulomb disorder, which depends on the $D\left(T_{\mathrm{e}}\right)$ \cite{Hwang2009e}. Figures\;B.1(e)--(g) present the energy dependence of the momentum relaxation rate via charged impurities with $n_{\mathrm{i}}=1.0 \times 10^{12} \mathrm{cm}^{-2}$. 

	\begin{figure}
		\centering
		\includegraphics[width=8.6cm, bb=0 0 241 190]{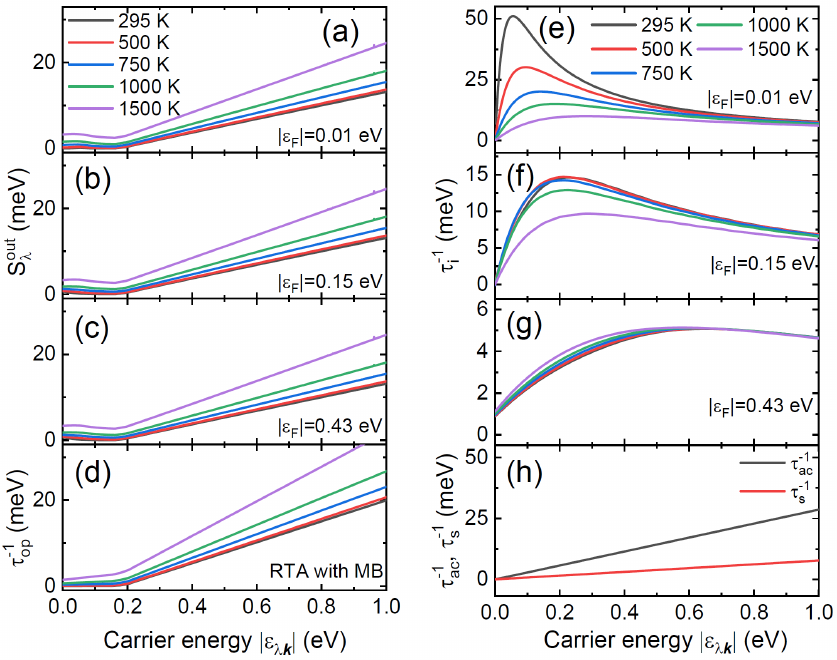}
		\caption{\label{figB1} $S_{\lambda}^{\text {out}}$ as a function of majority carrier energy $\varepsilon_{\lambda \textbf{\textit{k}}}=\lambda \hbar v_{\mathrm{F}}|\boldsymbol{k}| $ $(\lambda=\pm 1)$ for graphene with $\left|\varepsilon_{\mathrm{F}}\right|=$ (a) 0.01, (b) 0.15, and (c) $0.43\,\mathrm{eV}$. (d) Relaxation rate $\tau_{\mathrm{op}}^{-1}$ of optical phonons calculated by RTA assuming the Maxwell--Boltzmann distribution instead of the FD distribution. Relaxation rate $\tau_{\mathrm{i}}^{-1}$ of charged impurity with $n_{\mathrm{i}}=1.0 \times 10^{12}\,\mathrm{cm}^{-2}$ for $\left|\varepsilon_{\mathrm{F}}\right|=$ (e) 0.01, (f) 0.15, (g) $0.43\,\mathrm{eV}$, and (h) $\tau_{\text {ac }}^{-1}$ of acoustic phonons for $D_{\text {ac }}=30\,\mathrm{eV}$ (black line) and $\tau_{\mathrm{s}}^{-1}$ of weak scatterers for $\rho_{\mathrm{s}}=100\,\Omega$ (red line) at $T=295\,\mathrm{K}$. The parameters used in the calculation are summarized in Table I, and these results are also valid for the minority carriers except for  $S_{\lambda}^{\text {out}}$.}
	\end{figure}
	
	The possible physical origins of weak scatterers are ripples and point defects. The relaxation rate, $\tau_{\mathrm{s}}^{-1}\left(\varepsilon_{\lambda \textbf{\textit{k}}}\right)$, for the weak scatterers is expressed by \cite{Lin2014d}
	\begin{equation}
	\tau_{\mathrm{s}}^{-1}\left(\varepsilon_{\lambda \textbf{\textit{k}}}\right)=\frac{e^{2}}{\pi \hbar^{2}} \rho_{\mathrm{s}}\left|\varepsilon_{\lambda \textbf{\textit{k}}}\right|,
	\end{equation}
	where $\rho_{\mathrm{s}}$ is the resistivity of the weak scatterers and ranges from 40--100 $\Omega$ \cite{Morozov2008b, Dean2010b, Pachoud2010b, Yan2011b}. 
	
	The acoustic phonon scattering is treated as quasielastic, and the relaxation rate, $\tau_{\mathrm{ac}}^{-1}\left(\varepsilon_{\lambda \textbf{\textit{k}}}\right)$, can be expressed as cited in \cite{Perebeinos2010b}:
	\begin{equation}
	\tau_{\mathrm{ac}}^{-1}\left(\varepsilon_{\lambda \textbf{\textit{k}}}\right)=\frac{k_{\mathrm{B}} T_{\mathrm{e}}}{\hbar^{3} v_{\mathrm{F}}^{2}} \frac{D_{\mathrm{ac}}^{2}}{\rho v_{\mathrm{ph}}^{2}}\left|\varepsilon_{\lambda \textbf{\textit{k}}}\right|.
	\end{equation}
	Here, $D_{\text {ac}}$ is the acoustic deformation potential, which ranges from 10--$30\,\mathrm{eV}$ \cite{Chen2008f,Ono1966, Sugihara1983a, Bolotin2008b, Hong2009a}, and $v_{\mathrm{ph}}=2.0 \times 10^{4}\,\mathrm{m}\mathrm{s}^{-1}$ is the sound velocity in graphene. 
	In this study, we used $\rho_{\mathrm{s}}=100$ $\Omega$ and $D_{\mathrm{ac}}=30\,\mathrm{eV}$ in the calculation of the energy dependence of $\tau_{\mathrm{s}}^{-1}$ and $\tau_{\mathrm{ac}}^{-1}$, respectively, as illustrated in Fig.\;B.1(h). As a result, the elastic scattering term, $\nu^{\mathrm{el}}$, in Eq.\;(13) reads 
	\begin{equation}
			\nu^{\mathrm{el}}=\tau_{\mathrm{i}}^{-1}+\tau_{\mathrm{s}}^{-1}+\tau_{\mathrm{ac}}^{-1}.
	\end{equation}
	
	In the RTA, the collision term in Eq.\;(1) under the low field is expressed in the form 
	\begin{equation}
	\left.\frac{\partial f_{\lambda}(\boldsymbol{k}, t)}{\partial t}\right|_{\text {collision }}=\frac{f_{\lambda}(\boldsymbol{k}, t)-f_{0}\left(\varepsilon_{\lambda \textbf{\textit{k}}}\right)}{\tau\left(\varepsilon_{\lambda \textbf{\textit{k}}}\right)}.
	\end{equation}
	Here, $\tau^{-1}\left(\varepsilon_{\lambda \textbf{\textit{k}}}\right)$ is the relaxation rate of the distribution function. Using $\tau^{-1}\left(\varepsilon_{\lambda \textbf{\textit{k}}}\right)$, the intraband optical conductivity in graphene is given by \cite{Stauber2007b}
	\begin{equation}
	\sigma(\omega)=-\frac{e^{2} v_{F}}{2} \sum_{\lambda} \int_{-\infty}^{\infty} d \varepsilon_{\lambda \textbf{\textit{k}}} \frac{N\left(\varepsilon_{\lambda \textbf{\textit{k}}}\right)}{\tau^{-1}\left(\varepsilon_{\lambda \textbf{\textit{k}}}\right)-i \omega} \frac{\partial f_{0}\left(\varepsilon_{\lambda \textbf{\textit{k}}}\right)}{\partial \varepsilon_{\lambda \textbf{\textit{k}}}}.
	\end{equation}

\section{Calculation of temperature evolution of hot-carrier quasi-equilibrium state} 
The temperature model was used in previous studies \cite{Lui2010b, Lin2013d, Someya2017b, Wang2010a, Johannsen2013b, Sun2011a}. However, we modified the model to include the carrier energy relaxation (e.g., the SC carrier-cooling and optical phonon emission and absorption process). This can be described using the coupled rate equations of Eqs.\;(22)--(25). Moreover, we used the specific heat capacity, $C$, considering the temperature dependence of $\mu(T_{\mathrm{e}})$ of 2D-MDF as shown in Fig.A1(b). Here, $C=C_{c}+C_{v}$ is the sum of the specific heat of the electrons in the conduction bands and valence bands given by \cite{KittelC2004a}:
\begin{equation}
\begin{aligned}
C\left(T_{\mathrm{e}}\right) =&\frac{d U_{\mathrm{c}}}{d T_{\mathrm{e}}}+\frac{d U_{\mathrm{v}}}{d T_{\mathrm{e}}} \\
=&\int_{0}^{\infty} \varepsilon_{1 \textbf{\textit{k}}} N\left(\varepsilon_{1 \textbf{\textit{k}}}\right) \frac{d f_{0}\left(\varepsilon_{1 \textbf{\textit{k}}}\right)}{d T_{\mathrm{e}}} d \varepsilon_{1 \textbf{\textit{k}}} \\
&+\int_{-\infty}^{0} \varepsilon_{-1 \textbf{\textit{k}}} N\left(\varepsilon_{-1 \textbf{\textit{k}}}\right) \frac{d f_{0}\left(\varepsilon_{-1 \textbf{\textit{k}}}\right)}{d T_{\mathrm{e}}} d \varepsilon_{-1\textbf{\textit{k}}},
\end{aligned}
\end{equation}
where $U_{\mathrm{c}}$ and $U_{\mathrm{v}}$\ are the thermal kinetic energy of the electrons in the conduction and valence bands, respectively. 
Note that $C\left(T_{\mathrm{e}}\right), \varepsilon_{\lambda \textbf{\textit{k}}},$ and $\nu\left(\varepsilon_{\lambda \textbf{\textit{k}}}\right)$ are functions of the Fermi velocity, $v_{\mathrm{F}}$, which is renormalized by electron--electron, electron--phonon, and electron--plasmon coupling and depends on the carrier concentration owing to the carrier-screening effect in graphene \cite{Lazzeri2008b, Gonzalez1994a, Abedinpour2011a, Siegel2011a, Elias2011c, Kotov2012a, Astrakhantsev2015a, Stauber2017b, Park2007c, Pisana2007a, Stauber2008d, Basko2008e, Gonzalez1999, Carbotte2010c, LeBlanc2011a, Hwang2012e, Hwang2007i, Polini2008a, Bostwick2010b, Hwang2007g, DasSarma2007b, Bostwick2007b, Foster2008a, Trevisanutto2008a, Hwang2008e, Park2009d}. 
However, the screening constant of the electron--electron interaction has become a subject of considerable debate \cite{Barlas2007a, Hwang2007j, Fritz2008a, Kotov2008c, Reed2010b}. 
For simplicity, we used $v_{\mathrm{F}}=1.1 \times 10^{6}\,\mathrm{m} \mathrm{s}^{-1}$ in the numerical calculation. 
Figure\;C.1(a) plots the temperature dependence of $C\left(T_{\mathrm{e}}\right)$ for graphene with $|\varepsilon_{\mathrm{F}|}=0.01, 0.15,$ and $0.43\,\mathrm{eV}$. 
For comparison, Fig.\;C.1(a) plots $C\left(T_{\mathrm{e}}\right)=\alpha T_{\mathrm{e}}^{2}$, where $\alpha=16.3 \times 10^{-8}$ $\mathrm{nJK}^{-3} \mathrm{cm}^{-2}$ for $v_{\mathrm{F}}=1.0 \times 10^{6}$ $\mathrm{m} \mathrm{s}^{-1},$ calculated from Eq.\;(C1) for undoped graphene with a constant chemical potential. 
Here, $\alpha=16.3 \times 10^{-8}$ $\mathrm{nJK}^{-3} \mathrm{cm}^{-2}$ is twice the value of $\alpha=8.14 \times 10^{-8}$ $ \mathrm{n} \mathrm{JK}^{-3} \mathrm{cm}^{-2}$, which only considers the electron-heat capacity in the conduction band, as reported in Ref.\,\cite{Sun2011a}. 
As can be observed, $C\left(T_{\mathrm{e}}\right)$ for graphene having $|\varepsilon_{\mathrm{F}}|=0.01, 0.15$ and $0.43\,\mathrm{eV}$ exhibits different behaviors from a quadratic dependence below $T_{\mathrm{e}}=2,000\,\mathrm{K}$. These deviations are attributed not only to the finite Fermi energy but also to the temperature-dependent chemical potential $\mu\left(T_{\mathrm{e}}\right)$. 

$R_{\eta}^{\mathrm{Net}}=R_{\eta}-G_{\eta}$ in Eq.\;(22) denotes the total balance between the optical phonon emission and absorption rate.
\begin{equation}
\begin{aligned}
R_{\eta}=& \frac{\left\langle D_{\eta}^{2}\right\rangle_{\mathrm{F}}}{\pi \rho \omega_{\eta} \hbar^{4} v_{\mathrm{F}}^{4}} \int_{-\infty}^{\infty} d \varepsilon_{\lambda \textbf{\textit{k}}}\left|\varepsilon_{\lambda \textbf{\textit{k}}}\right|\left|\varepsilon_{\lambda \textbf{\textit{k}}}-\hbar \omega_{\eta}\right| \\
& \times f_{0}\left(\varepsilon_{\lambda \textbf{\textit{k}}}\right)\left(1-f_{0}\left(\varepsilon_{\lambda \textbf{\textit{k}}}-\hbar \omega_{\eta}\right)\right)\left(1+n_{\eta}\right),
\end{aligned}
\end{equation}
\begin{equation}
\begin{aligned}
G_{\eta} &=\frac{\left\langle D_{\eta}^{2}\right\rangle_{\mathrm{F}}}{\pi \rho \omega_{\eta} \hbar^{4} v_{\mathrm{F}}^{4}} \int_{-\infty}^{\infty} d \varepsilon_{\lambda \textbf{\textit{k}}}\left|\varepsilon_{\lambda \textbf{\textit{k}}}\right|\left|\varepsilon_{\lambda \textbf{\textit{k}}}+\hbar \omega_{\eta}\right| \\
& \times f_{0}\left(\varepsilon_{\lambda \textbf{\textit{k}}}\right)\left(1-f_{0}\left(\varepsilon_{\lambda \textbf{\textit{k}}}+\hbar \omega_{\eta}\right)\right) n_{\eta}.
\end{aligned}
\end{equation}
Figure\;C.2 compares the $T_{\mathrm{e}}$ dependence of the energy exchange rates, $R_{\eta} \hbar \omega_{\eta}$ and $G_{\eta} \hbar \omega_{\eta}$, for optical phonon emission and absorption in graphene at different Fermi energies and phonon temperatures, $T_{\eta}$. For $T_{\eta}<T_{\mathrm{e}}$, $R_{\eta} \hbar \omega_{\eta}$ is larger than $G_{\eta} \hbar \omega_{\eta}$, as indicated in Figs.\;C.2(a)--(c); the hot-carrier energy is transferred to the optical phonons. The total balance, $R_{\eta}^{\mathrm{Net}} \hbar \omega_{\eta}=R_{\eta} \hbar \omega_{\eta}-G_{\eta} \hbar \omega_{\eta}$, becomes zero when $T_{\eta}$ is equal to $T_{\mathrm{e}}$, as illustrated in Figs.\;C.2(d)--(f). However, the energy relaxation of the carriers and phonons is further driven by the optical-phonon decay rate, $\tau_{\mathrm{ph}}^{-1}$, caused by the anharmonic phonon--phonon interaction and SC process. 

\begin{figure}
	\centering
	\includegraphics[width=8.6cm, bb=0 0 241 93]{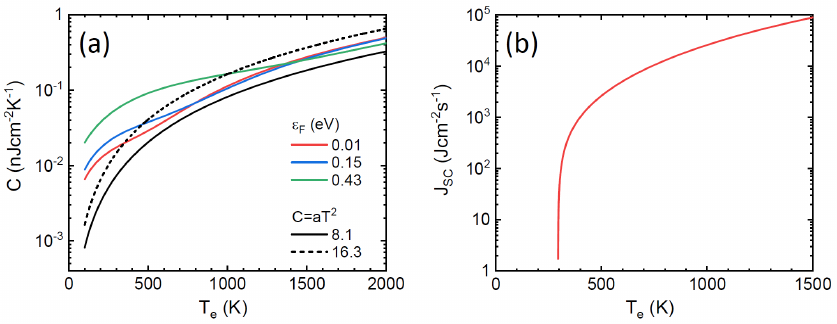}
	\caption{\label{figC1} (a) $T_{\mathrm{e}}$ dependence of specific-heat capacity of graphene with $|\varepsilon_{\mathrm{F}}|=0.01,$ (red line), 0.15 (blue line), and 0.43 (green line) eV calculated by considering temperature dependence of the chemical potential. For comparison, the black solid and broken lines indicate the specific-heat capacity for electrons in the conduction band and in both the conduction and valence bands, respectively, of undoped graphene with $\mu(T_{\mathrm{e}})=0$ eV. (b) The $T_{\mathrm{e}}$ dependence on $J_{\mathrm{sc}}$ for the SC process is calculated with $\mu_{\mathrm{m}}=4,800\,\mathrm{cm}^{-2} \mathrm{V}^{-1} \mathrm{s}^{-1}$ for scattering by short-range weak scatterers.}
\end{figure}

\begin{figure}
	\centering
	\includegraphics[width=8.6cm, bb=0 0 241 166]{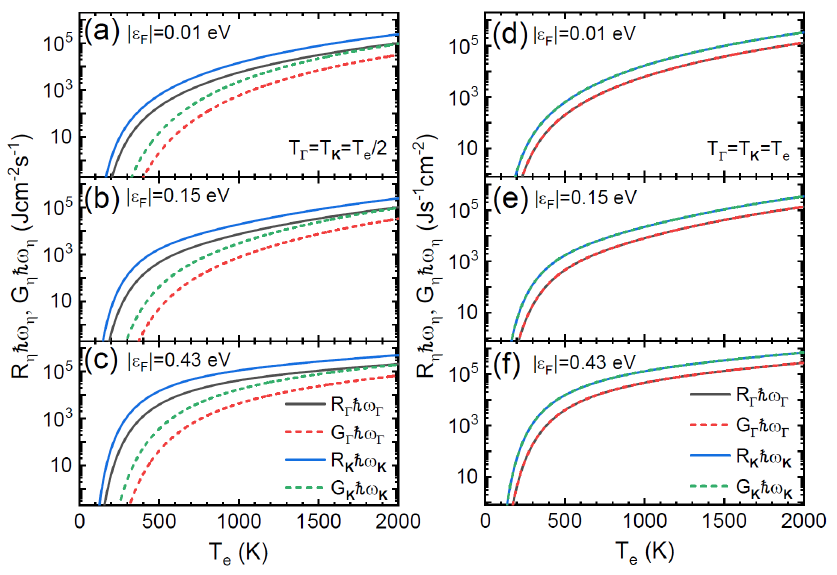}
	\caption{\label{figC2} $T_{\mathrm{e}}$ dependence of $R_{\eta} \hbar \omega_{\eta}$ and $G_{\eta} \hbar \omega_{\eta}$ for optical phonon emission and absorption processes, respectively, in graphene with $\left|\varepsilon_{\mathrm{F}}\right|=$ (a) 0.01, (b) 0.15, and (c) $0.43\,\mathrm{eV}$ for $T_{\eta}=T_{\mathrm{e}} / 2$ and those with $\left|\varepsilon_{\mathrm{F}}\right|=$(d) 0.01, (e) 0.15, and (f) $0.43\,\mathrm{eV}$ for $T_{\eta}=T_{\mathrm{e}}$.}
\end{figure}

\begin{figure}
	\centering
	\includegraphics[width=8.6cm, bb=0 0 283 190]{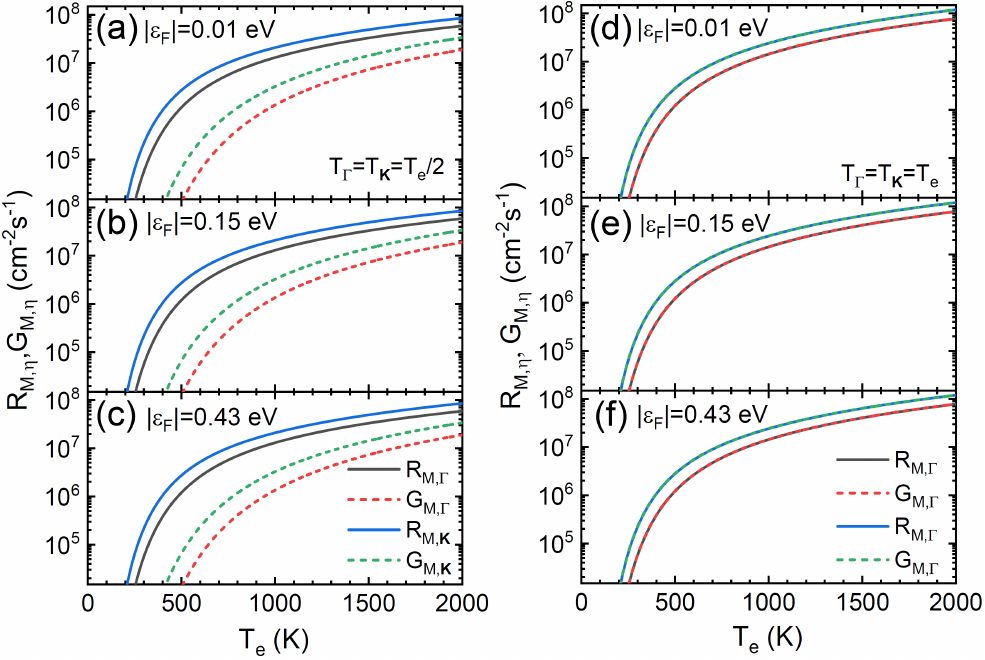}
	\caption{\label{figC3} $T_{\text {e }}$ dependence of $R_{\mathrm{M}, \eta}$ and $G_{\mathrm{M}, \eta}$ for optical phonon emission and absorption processes, respectively, in graphene with $\left|\varepsilon_{\mathrm{F}}\right|=$ (a) 0.01, (b) 0.15, and (c) $0.43\,\mathrm{eV}$ for $T_{\eta}=T_{\mathrm{e}} / 2$ and those with $\left|\varepsilon_{\mathrm{F}}\right|=$ (d) 0.01, (e) 0.15, and (f) $0.43\,\mathrm{eV}$ for $T_{\eta}=T_{\mathrm{e}}$.}
\end{figure}

$J_{\mathrm{sc}}$ in Eq.\;(22) denotes the energy loss rate for the SC process, which is a disorder-mediated electron--acoustic phonon scattering that takes place via a three-body collision involving a carrier, a defect, and an acoustic phonon. Although energy relaxation via acoustic phonon scattering is essential for low-energy carriers, the small Fermi surface and momentum conservation severely constrain the phase space of the acoustic phonon-scattering process. As a result, acoustic phonon scattering becomes an inefficient cooling channel. However, additional momentum exchanges with disorders enable acoustic phonons to use a much wider phase space, thereby enabling a larger dissipation of energy from the hot carriers. Thus, SC becomes an efficient cooling pathway. 
According to Ref.\,\cite{Song2012a}, $J_{\mathrm{sc}}$ is given by
\begin{equation}
J_{\mathrm{sc}}=\frac{9.62 g_{a c}^{2} N^{\prime 2}\left(\varepsilon_{\mathrm{F}}\right) k_{B}^{3}} {\hbar k_{F} l}\left(T_{\mathrm{e}}^{3}-T_{\mathrm{ac}}^{3}\right)\quad\left(\mathrm{eVs}^{-1}\right),
\end{equation}
where $g_{\mathrm{ac}}=D_{\mathrm{ac}} / 2 \rho v_{\mathrm{ph}}^{2}$ is the electron--acoustic phonon coupling constant, $N^{\prime}\left(\varepsilon_{F}\right)$ is the density of states at the Fermi energy per one spin or valley flavor, $k_{\mathrm{F}}$ is the Fermi wave number, $l$ is the mean free path of the short-range weak scatterers, and $T_{\text {ac }}$ is the acoustic phonon temperature, which is assumed to remain unchanged from the equilibrium state. Furthermore, $J_{\mathrm{sc}}$ can be expressed as a function of the carrier mobility, $\mu_{\mathrm{m}}=\sigma_{\mathrm{DC}} / n_{c} e$, for short-range weak scatterers, as discussed in Ref.\,\cite{Someya2017b}:
\begin{equation}
J_{\mathrm{sc}}\approx8.8 \times 10^{14} \frac{D_{a c}^{2}}{\mu_{\mathrm{m}}}\left(T_{\mathrm{e}}^{3}-T_{\mathrm{ac}}^{3}\right) \quad\left(\mathrm{eVs}^{-1}\right).
\end{equation}
For simplicity, we used the carrier mobility, $\mu_{\mathrm{m}}=4,800$ $\mathrm{cm}^{2} \mathrm{V}^{-1} \mathrm{s}^{-1}$, for SC-carrier cooling throughout the calculation. The $T_{\mathrm{e}}$ dependence of $J_{\mathrm{sc}}$ is plotted in Fig.\;C. 1(b).

In Eqs.\;(23)--(25), $R_{\mathrm{M}, \mathrm{n}}^{\mathrm{Net}}=R_{\mathrm{M}, \mathrm{\eta}}-G_{\mathrm{M}, \mathrm{\eta}}$ denotes the total balance between the optical-phonon emission and absorption rate per number of phonon modes.
\begin{equation}
\begin{aligned}
R_{\mathrm{M}, \eta} =&\frac{\left\langle D_{\eta}^{2}\right\rangle_{\mathrm{F}}}{\pi \rho \omega_{\eta} \hbar^{4} v_{\mathrm{F}}^{4}} \int_{-\infty}^{\infty} d \varepsilon_{\lambda \textbf{\textit{k}}}\left[\left|\varepsilon_{\lambda \textbf{\textit{k}}}\right|\left|\varepsilon_{\lambda \textbf{\textit{k}}}-\hbar \omega_{\eta}\right|\right.f_{0}\left(\varepsilon_{\lambda \textbf{\textit{k}}}\right)\\
&\left.\times \left(1-f_{0}\left(\varepsilon_{\lambda \textbf{\textit{k}}}-\hbar \omega_{\eta}\right)\right)\left(1+n_{\eta}\right)\right] / M_{\eta}^{-}\left(\varepsilon_{\lambda \textbf{\textit{k}}}\right),
\end{aligned}
\end{equation}
\begin{equation}
\begin{aligned}
G_{\mathrm{M}, \mathrm{\eta}} =&\frac{\left\langle D_{\mathrm{\eta}}^{2}\right\rangle_{\mathrm{F}}}{\pi \rho \omega_{\mathrm{\eta}} \hbar^{4} v_{\mathrm{F}}^{4}} \int_{-\infty}^{\infty} d \varepsilon_{\lambda \textbf{\textit{k}}}\left[\left|\varepsilon_{\lambda \textbf{\textit{k}}}\right|\left|\varepsilon_{\lambda \textbf{\textit{k}}}+\hbar \omega_{\mathrm{n}}\right|\right.f_{0}\left(\varepsilon_{\lambda \textbf{\textit{k}}}\right)\\ &\left.\times \left(1-f_{0}\left(\varepsilon_{\lambda \textbf{\textit{k}}}+\hbar \omega_{\mathrm{n}}\right)\right) n_{\mathrm{\eta}}\right] / M_{\mathrm{\eta}}^{+}\left(\varepsilon_{\lambda \textbf{\textit{k}}}\right).
\end{aligned}
\end{equation}
In the above equation, $M_{\eta}^{-}\left(\varepsilon_{\lambda \textbf{\textit{k}}}\right)$ and $M_{\eta}^{+}\left(\varepsilon_{\lambda \textbf{\textit{k}}}\right)$ are the numbers of $\eta$ phonon modes per unit area that participate in the carrier--phonon scattering of the emission and absorption processes for carriers having energy, $\varepsilon_{\lambda \textbf{\textit{k}}}$, respectively.
\begin{equation}
M_{\eta}^{\pm}\left(\varepsilon_{\lambda \textbf{\textit{k}}}\right)=\left|\frac{a_{\eta}}{4 \pi}\left\{\left(\frac{2 \varepsilon_{\lambda \textbf{\textit{k}}} \pm \hbar \omega_{\eta}}{\hbar v_{\mathrm{F}}}\right)^{2}-\left(\frac{\omega_{\eta}}{v_{\mathrm{F}}}\right)^{2}\right\}\right|.
\end{equation}
In this case, $a_{\Gamma}=1$ for $\Gamma\mathchar`-\mathrm{LO}$ and $\Gamma\mathchar`-\mathrm{TO}$ phonons, and $a_{\textbf{K}}=2$ for \textbf{K} phonons. The factor of $a_{\textbf{K}}=2$ for the \textbf{K} phonons represents the degenerate phonon valleys at the \textbf{K} and \textbf{K}’ points. Fig.\;(C.3) compares the $T_{\mathrm{e}}$ dependence of the energy-exchange rates, $R_{\mathrm{M}, \mathrm{\eta}}$ and $G_{\mathrm{M}, \mathrm{\eta}}$, for the optical phonon emission and absorption in graphene at different Fermi energies and optical phonon temperatures, $T_{\eta},$ that demonstrate similar behaviors to those of $R_{\eta} \hbar \omega_{\eta}$ and $G_{\eta} \hbar \omega_{\eta}$.

\section{Calculation of transmission change of THz-probe pulses}
When applying the standard thin-film approximation \cite{Bludov2013a}, the THz-amplitude transmission coefficient of monolayer graphene having complex conductivity, $\sigma(\omega)$, on a substrate with a thickness, $d$, at normal incidence is expressed by the ratio of the incident wave, $E_{\mathrm{i}}(\omega)$, and the transmitted wave, $E_{\mathrm{t}}(\omega)$.
\begin{equation}
\begin{aligned}
t(\omega) &=\frac{E_{\mathrm{t}}(\omega)}{E_{\mathrm{i}}(\omega)} \\
&=\frac{2 \epsilon_{0}^{1 / 2}}{\sigma(\omega) Z_{0}+\epsilon_{\mathrm{THz}}^{1 / 2}+\epsilon_{v}^{1 / 2}} \frac{2 \epsilon_{\mathrm{THz}}^{1 / 2}}{\epsilon_{\mathrm{THz}}^{1 / 2}+\epsilon_{v}^{1 / 2}} e^{-i \epsilon_{\mathrm{THz}}^{1/2} d \omega / c}.
\end{aligned}
\end{equation}
Here, $\epsilon_{\mathrm{THz}}$ and $\epsilon_{v}$ are the dielectric constant of the substrate at THz frequency range and vacuum, $Z_{0}$ is the vacuum impedance, and $c$ is the speed of light in a vacuum. Similarly, the THz-amplitude transmission coefficient of monolayer graphene under photoexcitation with a pump-probe delay, $\tau_{1}$, having a complex conductivity, $\sigma\left(\omega, \tau_{1}\right)$, is given by
\begin{equation}
\begin{aligned}
t\left(\omega, \tau_{1}\right) =&\frac{E_{t}\left(\omega, \tau_{1}\right)}{E_{i}(\omega)} \\
=&\frac{2 \epsilon_{0}^{1 / 2}}{\sigma\left(\omega, \tau_{1}\right) Z_{0}+\epsilon_{\mathrm{THz}}^{1 / 2}+\epsilon_{v}^{1 / 2}}\\
&\times \frac{2 \epsilon_{\mathrm{THz}}^{1 / 2}}{\epsilon_{\mathrm{THz}}^{1 / 2}+\epsilon_{v}^{1 / 2}} e^{-i \epsilon_{\mathrm{THz}}^{1 / 2} d \omega / c},
\end{aligned}
\end{equation}
where $E_{\mathrm{t}}\left(\omega, \tau_{1}\right)$ is the transmitted THz electric field through the photoexcited graphene at $\tau_{1}$. According to Eqs.\;(D1) and (D2), the frequency-dependent complex photoconductivity, $\Delta \sigma\left(\omega, \tau_{1}\right)=\sigma\left(\omega, \tau_{1}\right)-\sigma_{0}(\omega)$, where $\sigma_{0}(\omega)$ is the complex conductivity without pump fluence again, is expressed by
\begin{equation}
\Delta \sigma\left(\omega, \tau_{1}\right)=-\left(\frac{\sigma\left(\omega, \tau_{1}\right) Z_{0}+\epsilon_{\mathrm{THz}}^{1 / 2}+\epsilon_{v}^{1 / 2}}{Z_{0}}\right) \frac{\Delta E_{t}\left(\omega, \tau_{1}\right)}{E_{t}(\omega)}.
\end{equation}
Here, $\Delta E_{\mathrm{t}}\left(\omega, \tau_{1}\right)=E_{\mathrm{t}}\left(\omega, \tau_{1}\right)-E_{t}(\omega)$ is the change in the THz electric field in the frequency domain. 
Subsequently, the transmitted THz field in the time domain, $E_{t}\left(\tau_{2}, \tau_{1}\right)$, where $\tau_{2}$ is the probe-trigger delay, is given by
\begin{equation}
\begin{aligned}
E_{\mathrm{t}}\left(\tau_{2}, \tau_{1}\right) &=\int E_{\mathrm{t}}\left(\omega, \tau_{1}\right) e^{i \omega \tau_{2}} d \omega \\
&=\int E_{\mathrm{i}}\left(\omega_{1}\right) t\left(\omega, \tau_{1}\right) e^{i \omega \tau_{2}} d \omega.
\end{aligned}
\end{equation}
The normalized negative transmission change, $ -\Delta E_{\mathrm{t}}\left(\tau_{2}, \tau_{1}\right)/E_{t}\left(\tau_{2}\right) $, as a function of the probe-trigger delay, $\tau_{2}$, at $\tau_{1}$ is expressed by
\begin{equation}
-\frac{\Delta E_{\mathrm{t}}\left(\tau_{2}, \tau_{1}\right)}{E_{t}\left(\tau_{2}\right)}=-\frac{E_{\mathrm{t}}\left(\tau_{2}, \tau_{1}\right)-E_{t}\left(\tau_{2}\right)}{E_{t}\left(\tau_{2}\right)}.
\end{equation}
Here, $E_{\mathrm{t}}\left(\tau_{2}\right)$ is the transmitted THz field through the graphene sample without pump fluence. Figure\;D.1 plots $\Delta E_{t}\left(\tau_{1}\right) / E_{0} \equiv \Delta E_{t}\left(0, \tau_{1}\right) / E_{t}(0)$ of the undoped graphene with $\varepsilon_{\mathrm{F}}=-0.01\,\mathrm{eV}$ and $n_{\mathrm{iH}}$, calculated using the THz probe with pulse durations of $2 \tau_{\mathrm{p}}=300$ and $500\,\mathrm{fs}$. Their corresponding central frequencies are $\omega / 2\pi=2.2$ and $1.2\,\mathrm{THz}$, respectively. In this case, $-\Delta E_{\mathrm{t}}\left(\tau_{1}\right) / E_{0}$ for $2 \tau_{\mathrm{p}}=500\,\mathrm{fs}$ exhibits a faster decay time than that of $2 \tau_{\mathrm{p}}=300\,\mathrm{fs}$, which indicates the difference of the decay time of $\Delta \sigma\left(\omega, \tau_{1}\right)$ between 1.2 and $2.2\,\mathrm{THz}$.
\begin{figure}[t]
	\centering
	\includegraphics[width=6cm, bb=0 0 283 226]{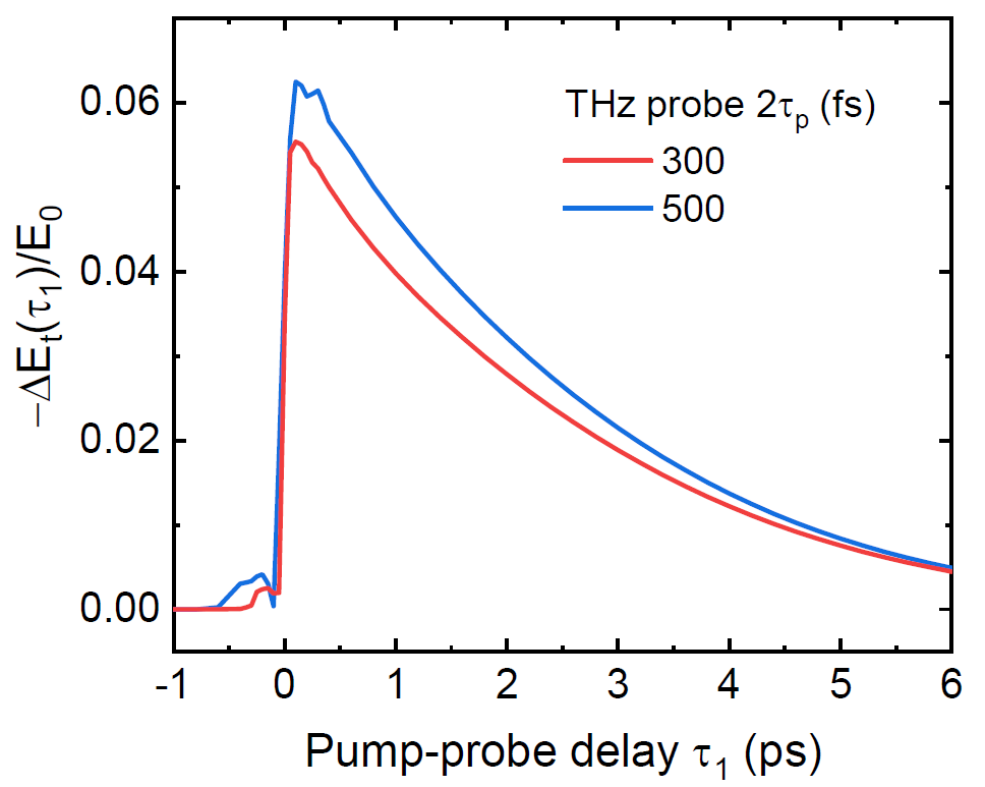}
	\caption{\label{figD1} $-\Delta E_{\mathrm{t}}\left(\tau_{1}\right) / E_{0}$ of undoped graphene with $n_{\mathrm{iH}}=1.0 \times 10^{12}\,\mathrm{cm}^{-2}$ for the THz probe with the pulse durations of $2 \tau_{\mathrm{p}}=300\,\mathrm{fs}$ (red line) and $500\,\mathrm{fs}$ (blue line).}
\end{figure}

\section{Convergence of electrical and optical conductivity via iterative calculation} 
The convergence of the iterative calculation, including inelastic scattering, is presented. Figure\;E.1(a) shows the dependence of the DC conductivity, $\sigma_{\mathrm{DC}}$, in the highly doped graphene under a constant electric field, $100\,\mathrm{Vcm^{-1}}$, on $j$ iterations using Eq.\;(14) considering only optical phonon scattering at different temperatures. 
At $T_{\mathrm{e} }=295\,\mathrm{K}$, $\sigma_{\mathrm{DC}}$ converged very rapidly, and even at $j=1$. The error, $|\sigma_{\mathrm{DC}}^{\mathrm{c}}-\sigma_{\mathrm{DC}}^{j}| / \sigma_{\mathrm{DC}}^{\mathrm{c}}<10^{-2},$ where $\sigma_{\mathrm{DC}}^{\mathrm{c}}$ and $\sigma_{\mathrm{DC}}^{j}$ are the converged DC conductivity and that at the $j$th iteration, respectively. 
However, the convergence rate became slower, and the error of the first iteration increased as the temperature increased, which can be attributed to the broader carrier distribution and increased scattering rate by optical phonons at higher temperatures. 
Figure\;E.1(b) presents the converged $g\left(\varepsilon_{\lambda \textbf{\textit{k}}}\right)$ of the heavily doped graphene under an electric field of $100\,\mathrm{Vcm}^{-1}$ calculated by Eq.\;(14). The $g\left(\varepsilon_{\lambda \textbf{\textit{k}}}\right)$ value showed a clear peak around $\varepsilon_{\mathrm{F}}=-0.43\,\mathrm{eV}$ at $T_{\mathrm{e}}=295\,\mathrm{K}$ spread and decreased with an increasing temperature that resulted in a reduction in $\sigma_{\mathrm{DC}}^{\mathrm{c}}$ at high temperatures.
\begin{figure}
	\centering
	\includegraphics[width=8.6cm, bb=0 0 283 106]{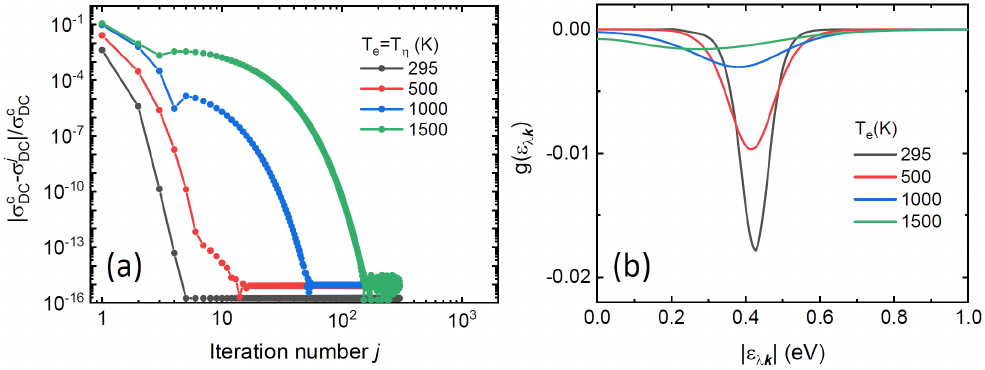}
	\caption{\label{figE1} (a) Dependence of error $\left|\sigma_{\mathrm{DC}}^{c}-\sigma_{\mathrm{DC}}^{j}\right| / \sigma_{\mathrm{DC}}^{c}$ on the number of iterations calculated by considering optical phonon scattering at $T_{\mathrm{e}}=295, 500, 1,000$, and $1,500\,\mathrm{K}$. (b) Converged $g\left(\varepsilon_{\lambda \textbf{\textit{k}}}\right)$ caused by the DC electric field of $100\,\mathrm{Vcm}^{-1}$ used for the calculation of $\sigma_{\mathrm{DC}}^{c}$.}
\end{figure}
\begin{figure}
	\centering
	\includegraphics[width=8.6cm, bb=0 0 283 113]{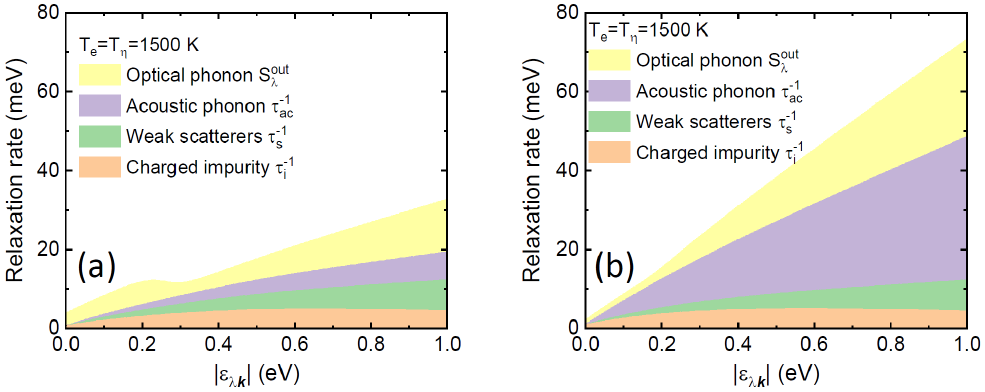}
	\caption{\label{figE2} Carrier-energy dependence of $S^{\mathrm{out}}_{\lambda}$ and momentum relaxation rates for acoustic phonon, weak scatterers, and charged impurity scattering at $T_{\mathrm{e}}=$ (a) $295\,\mathrm{K}$ and (b) $1500\,\mathrm{K}$.}
\end{figure}
\begin{figure}[b]
	\centering
	\includegraphics[width=8.6cm, bb=0 0 283 207]{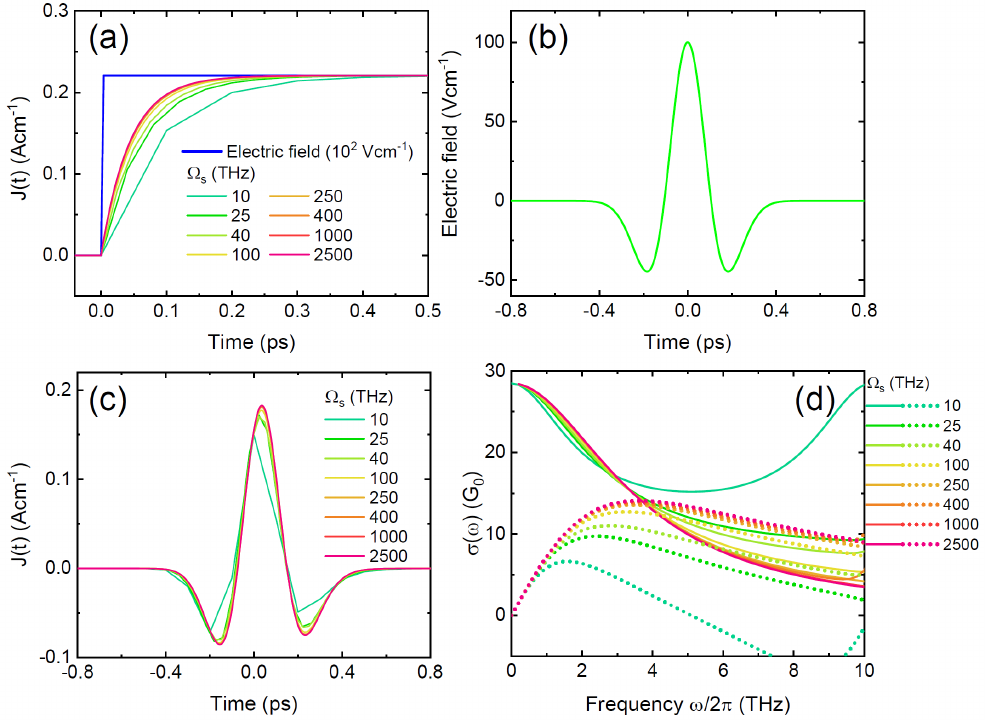}
	\caption{\label{figE3} (a) Temporal evolution of current in graphene at $T_{\mathrm{e}}=295\,\mathrm{K}$ driven by an electric field having a step function shape (blue line) calculated with $\Omega_{\mathrm{s}}=10,25,40,100,250,400,1,000,$ and $2,500\,\mathrm{THz}$. (b) Temporal waveform of the electric field of the THz-probe pulse. (c) THz-field induced intraband current generated in doped graphene and (d) the corresponding complex optical conductivity calculated with different $\Omega_{\mathrm{s}}$.}
\end{figure}

Moreover, we determined the convergence of the iterative solution for the time-dependent process given by Eq.\;(19) in which we considered the inelastic and elastic scatterings by optical phonons, acoustic phonons, charged impurities, and weak scatterers for  highly doped graphene. 
The momentum relaxation rates and $S^{\mathrm{out}}_{\lambda}$ used in the calculation at $T_{\mathrm{e}}=295$ and $1,500\,\mathrm{K}$ are illustrated in Fig.\;E.2. 
Fig.\;E.3(a) depicts the temporal evolution of the current density, $J(t)$, when an electric field having a step function shape was applied. The rising time of $J(t)$ decreased with an increasing $\Omega_{\mathrm{s}}$ in Eq.\;(19) and converged for $\Omega_{\mathrm{s}}>250\,\mathrm{THz},$ which was more than 10 times the momentum relaxation rate (approximately $25\,\mathrm{THz}$) at $\left|\varepsilon_{\lambda \textbf{\textit{k}}}\right|=\left|\varepsilon_{\mathrm{F}}\right|=0.43\,\mathrm{ eV}$. The $J(t)$ value converged to $0.22\,\mathrm{Acm}^{-1}$, yielding the DC conductivity, $\sigma_{\mathrm{DC}}^{\mathrm{c}}=28.5\mathrm{G}_{0}.$ Furthermore, we calculated the temporal variation in the current density induced by applying the THz-probe pulse with $2 \tau_{p}=300\,\mathrm{fs}$ as indicated in Fig.\;E.3(b). The temporal waveforms of $J(t)$ and the corresponding $\sigma(\omega)$ are plotted in Figs.\;E.3(c) and (d), respectively. 
Whereas the temporal waveform of $J(t)$ appears to be converged around $\Omega_{s}=250\,\mathrm{THz}$, the convergence of $\sigma(\omega)$ is strongly dependent on the frequency. Below $\omega / 2 \pi=0.2\,\mathrm{THz}$, the convergence is achieved even at $\Omega_{s}=10\,\mathrm{THz}$, and the convergence value of the DC conductivity of $\sigma(\omega)=28.5\mathrm{G}_{0}$ is equal to $\sigma_{\mathrm{DC}}^{\mathrm{c}}$, as obtained from the calculation by the step function-type electric field presented in Fig. E.3(a). However, $\sigma(\omega)$ at a higher frequency requires a larger $\Omega_{s}$ to be converged. In the numerical calculations explained in the main manuscript, we set $\Omega_{\mathrm{s}}=200$ and $1,000\,\mathrm{THz}$ for $\Delta E_{t}\left(\tau_{1}\right) / E_{0}$ and $\Delta \sigma\left(\omega, \tau_{1}\right)$, respectively. 

\bibliographystyle{apsrev4-2.bst}
\bibliography{library}

\end{document}